\pdfoutput=1
\documentclass[12pt,a4paper]{article}
\usepackage{ifthen}
\newboolean{pdflatex}
\setboolean{pdflatex}{true}

\newboolean{articletitles}
\setboolean{articletitles}{true}

\newboolean{uprightparticles}
\setboolean{uprightparticles}{false}

\newboolean{inbibliography}
\setboolean{inbibliography}{false}

\usepackage{microtype}
\usepackage{lineno}  % for line numbering during review
\usepackage{xspace} % To avoid problems with missing or double spaces after
\usepackage{caption} %these three command get the figure and table captions automatically small

\usepackage{graphicx}  % to include figures (can also use other packages)
\usepackage{color}
\usepackage{colortbl}
\graphicspath{{./figs/}} % Make Latex search fig subdir for figures

\usepackage{amsmath} % Adds a large collection of math symbols
\usepackage{amssymb}
\usepackage{amsfonts}
\usepackage{upgreek} % Adds in support for greek letters in roman typeset

\newcommand*\patchAmsMathEnvironmentForLineno[1]{%
\expandafter\let\csname old#1\expandafter\endcsname\csname #1\endcsname
\expandafter\let\csname oldend#1\expandafter\endcsname\csname
end#1\endcsname
 \renewenvironment{#1}%
   {\linenomath\csname old#1\endcsname}%
   {\csname oldend#1\endcsname\endlinenomath}%
}
\newcommand*\patchBothAmsMathEnvironmentsForLineno[1]{%
  \patchAmsMathEnvironmentForLineno{#1}%
  \patchAmsMathEnvironmentForLineno{#1*}%
}
\AtBeginDocument{%
\patchBothAmsMathEnvironmentsForLineno{equation}%
\patchBothAmsMathEnvironmentsForLineno{align}%
\patchBothAmsMathEnvironmentsForLineno{flalign}%
\patchBothAmsMathEnvironmentsForLineno{alignat}%
\patchBothAmsMathEnvironmentsForLineno{gather}%
\patchBothAmsMathEnvironmentsForLineno{multline}%
\patchBothAmsMathEnvironmentsForLineno{eqnarray}%
}

\usepackage{hyperref}    % Hyperlinks in references
\usepackage[all]{hypcap} % Internal hyperlinks to floats.

\def\lhcb {\mbox{LHCb}\xspace}

\ifthenelse{\boolean{uprightparticles}}%
{

 \def\Pmu         {\ensuremath{\upmu}\xspace}

 \def\Ppi         {\ensuremath{\uppi}\xspace}

 \def\Ppsi        {\ensuremath{\uppsi}\xspace}

 \def\PDelta      {\ensuremath{\Delta}\xspace}                 
 \def\PXi      {\ensuremath{\Xi}\xspace}                 
 \def\PLambda      {\ensuremath{\Lambda}\xspace}                 
 \def\PSigma      {\ensuremath{\Sigma}\xspace}                 
 \def\POmega      {\ensuremath{\Omega}\xspace}                 
 \def\PUpsilon      {\ensuremath{\Upsilon}\xspace}                 
 
 \def\PB      {\ensuremath{\mathrm{B}}\xspace}                 
                  
 \def\PD      {\ensuremath{\mathrm{D}}\xspace}

 \def\PJ      {\ensuremath{\mathrm{J}}\xspace}                 
 \def\PK      {\ensuremath{\mathrm{K}}\xspace}

 \def\Pb      {\ensuremath{\mathrm{b}}\xspace}                 
 \def\Pc      {\ensuremath{\mathrm{c}}\xspace}

 \def\Pi      {\ensuremath{\mathrm{i}}\xspace}

 \def\Ps      {\ensuremath{\mathrm{s}}\xspace}

}
{

 \def\Pmu         {\ensuremath{\mu}\xspace}

 \def\Ppi         {\ensuremath{\pi}\xspace}

 \def\Ppsi        {\ensuremath{\psi}\xspace}                 
                  
 \mathchardef\PDelta="7101
 \mathchardef\PXi="7104
 \mathchardef\PLambda="7103
 \mathchardef\PSigma="7106
 \mathchardef\POmega="710A
 \mathchardef\PUpsilon="7107
                  
 \def\PB      {\ensuremath{B}\xspace}                 
                  
 \def\PD      {\ensuremath{D}\xspace}

 \def\PJ      {\ensuremath{J}\xspace}                 
 \def\PK      {\ensuremath{K}\xspace}

 \def\Pb      {\ensuremath{b}\xspace}                 
 \def\Pc      {\ensuremath{c}\xspace}

 \def\Pi      {\ensuremath{i}\xspace}

 \def\Ps      {\ensuremath{s}\xspace}

}

\makeatletter
\ifcase \@ptsize \relax% 10pt
  \newcommand{\miniscule}{\@setfontsize\miniscule{4}{5}}% \tiny: 5/6
\or% 11pt
  \newcommand{\miniscule}{\@setfontsize\miniscule{5}{6}}% \tiny: 6/7
\or% 12pt
  \newcommand{\miniscule}{\@setfontsize\miniscule{5}{6}}% \tiny: 6/7
\fi
\makeatother

\DeclareRobustCommand{\optbar}[1]{\shortstack{{\miniscule (\rule[.5ex]{1.25em}{.18mm})}
  \\ [-.7ex] $#1$}}

\def\mup        {{\ensuremath{\Pmu^+}}\xspace}
\def\squark    {{\ensuremath{\Ps}}\xspace}

\def\cquark    {{\ensuremath{\Pc}}\xspace}

\def\bquark    {{\ensuremath{\Pb}}\xspace}

\def\pion   {{\ensuremath{\Ppi}}\xspace}

\def\pip    {{\ensuremath{\pion^+}}\xspace}

  \def\Kbar    {{\kern 0.2em\overline{\kern -0.2em \PK}{}}\xspace}

\def\KorKbar    {\kern 0.18em\optbar{\kern -0.18em K}{}\xspace}

  \def\Dbar    {{\kern 0.2em\overline{\kern -0.2em \PD}{}}\xspace}

\def\DorDbar    {\kern 0.18em\optbar{\kern -0.18em D}{}\xspace}

\def\B       {{\ensuremath{\PB}}\xspace}
\def\Bbar    {{\ensuremath{\kern 0.18em\overline{\kern -0.18em \PB}{}}}\xspace}

\def\BorBbar    {\kern 0.18em\optbar{\kern -0.18em B}{}\xspace}

\def\Bu      {{\ensuremath{\B^+}}\xspace}

\def\Bs      {{\ensuremath{\B^0_\squark}}\xspace}

\def\Bc      {{\ensuremath{\B_\cquark^+}}\xspace}

\def\jpsi     {{\ensuremath{{\PJ\mskip -3mu/\mskip -2mu\Ppsi\mskip 2mu}}}\xspace}

  \def\Y#1S{\ensuremath{\PUpsilon{(#1S)}}\xspace}% no space before {...}!

\def\Lbar        {{\ensuremath{\kern 0.1em\overline{\kern -0.1em\PLambda}}}\xspace}
\def\LorLbar    {\kern 0.18em\optbar{\kern -0.18em \PLambda}{}\xspace}

\def\BF         {{\ensuremath{\cal B}}\xspace}

\def\BR         {\BF}
\def\to                 {\ensuremath{\rightarrow}\xspace}

\def\AT#1     {\ensuremath{A_{\mathrm{T}}^{#1}}\xspace}           % 2

\def\C#1      {\ensuremath{\mathcal{C}_{#1}}\xspace}                       % 9
\def\Cp#1     {\ensuremath{\mathcal{C}_{#1}^{'}}\xspace}                    % 7
\def\Ceff#1   {\ensuremath{\mathcal{C}_{#1}^{\mathrm{(eff)}}}\xspace}        % 9  
\def\Cpeff#1  {\ensuremath{\mathcal{C}_{#1}^{'\mathrm{(eff)}}}\xspace}       % 7
\def\Ope#1    {\ensuremath{\mathcal{O}_{#1}}\xspace}                       % 2
\def\Opep#1   {\ensuremath{\mathcal{O}_{#1}^{'}}\xspace}                    % 7

\newcommand{\tev}{\ifthenelse{\boolean{inbibliography}}{\ensuremath{~T\kern -0.05em eV}\xspace}{\ensuremath{\mathrm{\,Te\kern -0.1em V}}}\xspace}
\newcommand{\gev}{\ensuremath{\mathrm{\,Ge\kern -0.1em V}}\xspace}
\newcommand{\mev}{\ensuremath{\mathrm{\,Me\kern -0.1em V}}\xspace}
\newcommand{\kev}{\ensuremath{\mathrm{\,ke\kern -0.1em V}}\xspace}
\newcommand{\ev}{\ensuremath{\mathrm{\,e\kern -0.1em V}}\xspace}
\newcommand{\gevc}{\ensuremath{{\mathrm{\,Ge\kern -0.1em V\!/}c}}\xspace}
\newcommand{\mevc}{\ensuremath{{\mathrm{\,Me\kern -0.1em V\!/}c}}\xspace}
\newcommand{\gevcc}{\ensuremath{{\mathrm{\,Ge\kern -0.1em V\!/}c^2}}\xspace}
\newcommand{\gevgevcccc}{\ensuremath{{\mathrm{\,Ge\kern -0.1em V^2\!/}c^4}}\xspace}
\newcommand{\mevcc}{\ensuremath{{\mathrm{\,Me\kern -0.1em V\!/}c^2}}\xspace}

\def\mum  {\ensuremath{{\,\upmu\rm m}}\xspace}

\def\mub{\ensuremath{{\rm \,\upmu b}}\xspace}

\def\invfb   {\ensuremath{\mbox{\,fb}^{-1}}\xspace}

\def\gsim{{~\raise.15em\hbox{$>$}\kern-.85em
          \lower.35em\hbox{$\sim$}~}\xspace}
\def\lsim{{~\raise.15em\hbox{$<$}\kern-.85em
          \lower.35em\hbox{$\sim$}~}\xspace}

\def\ptot       {\mbox{$p$}\xspace}
\def\pt         {\mbox{$p_{\rm T}$}\xspace}

\def\bcvegpy    {\mbox{\textsc{Bcvegpy}}\xspace}

\def\evtgen     {\mbox{\textsc{EvtGen}}\xspace}

\def\geant      {\mbox{\textsc{Geant4}}\xspace}

\def\photos     {\mbox{\textsc{Photos}}\xspace}

\def\pythia     {\mbox{\textsc{Pythia}}\xspace}

\def\tell1  {TELL1\xspace}
\def\ukl1   {UKL1\xspace}

\newcommand{\ptrans}{\ensuremath{p_{\rm T}}}

\newcommand{\bsubc}{\ensuremath{B_c^+}}
\newcommand{\bplus}{\ensuremath{B^+}}
\newcommand{\bcjpsipi}{\ensuremath{\bsubc\to\jpsi \pi^+}}
\newcommand{\bcjpsik}{\ensuremath{\bsubc\to\jpsi K^+}}

\newcommand{\bpjpsik}{\ensuremath{\bplus\to\jpsi K^+}}
\newcommand{\bpjpsipi}{\ensuremath{\bplus\to\jpsi \pi^+}}
\newcommand{\jpsimumu}{\ensuremath{\jpsi\to\mu^+ \mu^-}}
\newcommand{\ratiotot}{\ensuremath{(0.683\,\pm\,0.018\,\pm\,0.009)\%}\xspace}

\usepackage{booktabs}
\usepackage{cite} % Allows for ranges in citations
\usepackage{mciteplus}

\begin{document}

\renewcommand{\thefootnote}{\fnsymbol{footnote}}
\setcounter{footnote}{1}

% $Id: title-LHCb-PAPER.tex 63239 2014-11-10 08:38:26Z bliu $
% ===============================================================================
% Purpose: LHCb-PAPER journal paper title page template
% Author:
% Created on: 2010-09-25
% ===============================================================================

%%%%%%%%%%%%%%%%%%%%%%%%%
%%%%%  TITLE PAGE  %%%%%%
%%%%%%%%%%%%%%%%%%%%%%%%%
\begin{titlepage}
\pagenumbering{roman}

% Header ---------------------------------------------------
\vspace*{-1.5cm}
\centerline{\large EUROPEAN ORGANIZATION FOR NUCLEAR RESEARCH (CERN)}
\vspace*{1.0cm}
\hspace*{-0.5cm}
\begin{tabular*}{\linewidth}{lc@{\extracolsep{\fill}}r}
\ifthenelse{\boolean{pdflatex}}% Logo format choice
{\vspace*{-2.7cm}\mbox{\!\!\!\includegraphics[width=.14\textwidth]{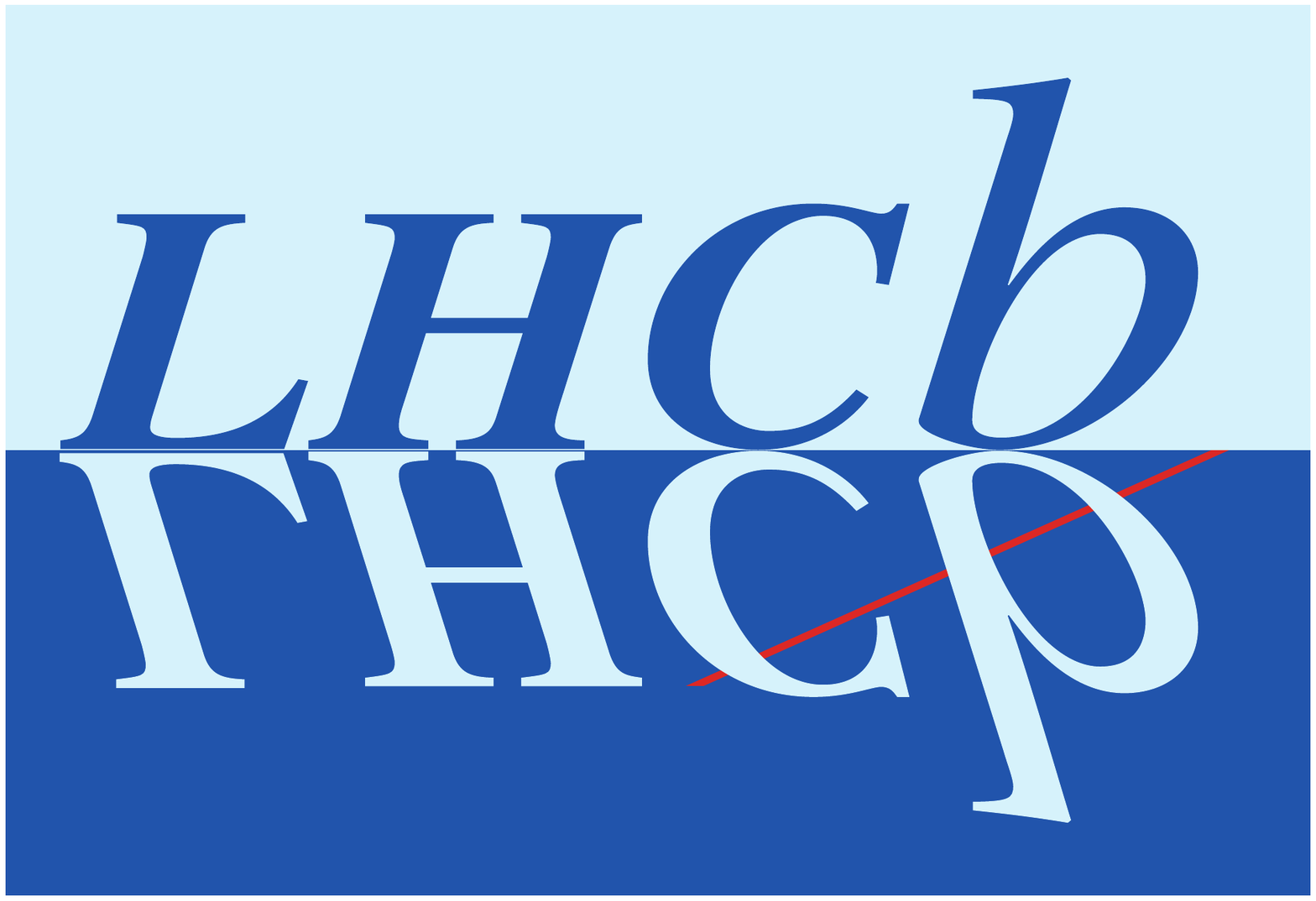}} & &}%
{\vspace*{-1.2cm}\mbox{\!\!\!\includegraphics[width=.12\textwidth]{lhcb-logo.eps}} & &}%
\\
 & & CERN-PH-EP-2014-269 \\  % ID
 & & LHCb-PAPER-2014-050 \\  % ID
 & & \today \\ % Date - Can also hardwire e.g.: 23 March 2010
 & & version 2.0\\
% not in paper \hline
\end{tabular*}

\vspace*{2.0cm}

% Title --------------------------------------------------
{\bf\boldmath\huge
\begin{center}
  Measurement of $B_c^+$ production in proton-proton collisions\\
  at $\sqrt{s}=8$ TeV
\end{center}
}

\vspace*{2.0cm}

% Authors -------------------------------------------------
\begin{center}
%In the footnote, replace 'paper' by 'letter' in case of submission to PRL or PLB
The LHCb collaboration\footnote{Authors are listed at the end of this Letter.}
\end{center}

\vspace{\fill}

% Abstract -----------------------------------------------
\begin{abstract}
  \noindent
  Production of $\Bc$ mesons in proton-proton collisions at
  a center-of-mass energy of 8~TeV is studied with data
  corresponding to an integrated luminosity
  of $2.0$\invfb recorded by the LHCb experiment.
  The ratio of production cross-sections times branching fractions between the
  \bcjpsipi\ and \bpjpsik\ decays
  is measured as a function of
  transverse momentum and rapidity in the regions
  $0 < \ptrans < 20\gevc$ and $2.0 < y < 4.5$.
  The ratio in this kinematic range is measured to be \ratiotot,
  where the first uncertainty is statistical and the second
  systematic.
\end{abstract}

\vspace*{2.0cm}

\begin{center}
  Published in Phys.~Rev.~Lett.\\
\end{center}

\vspace{\fill}

{\footnotesize
\centerline{\copyright~CERN on behalf of the \lhcb collaboration, licence \href{http://creativecommons.org/licenses/by/4.0/}{CC-BY-4.0}.}}
\vspace*{2mm}

\end{titlepage}

%%%%%%%%%%%%%%%%%%%%%%%%%%%%%%%%
%%%%%  EOD OF TITLE PAGE  %%%%%%
%%%%%%%%%%%%%%%%%%%%%%%%%%%%%%%%

%  empty page follows the title page ----
\newpage
\setcounter{page}{2}
\mbox{~}

\cleardoublepage

\renewcommand{\thefootnote}{\arabic{footnote}}
\setcounter{footnote}{0}

\pagestyle{plain}
\setcounter{page}{1}
\pagenumbering{arabic}

%\linenumbers

\noindent
In the Standard Model, the $B_c$ mesons
are the only states formed by two heavy quarks of different flavor,
the $\bar{b}$ and the $c$ quarks.
The production of $B_c$ mesons in hadron collisions
implies the simultaneous production of
$b\bar{b}$ and $c\bar{c}$ pairs, therefore it is rarer than that
of other $b$ mesons.
The production of $b\bar{b}$ and $c\bar{c}$ quarkonium states
in hadron collisions has been studied for two decades, however,
significant puzzles remain~\cite{Brambilla:2010cs}.
The relative role of competing production
mechanisms~\cite{Chang:1979nn,Baier:1983va,Bodwin:1994jh,Cho:1995vh,*Cho:1995ce}
is poorly understood and theory is unable to predict all
experimentally observed
features~\cite{Abe:1997jz,*Abe:1997yz,Abulencia:2007us,Abelev:2011md,Chatrchyan:2012woa,Aaij:2013nlm,Aaij:2014qea}.
The study of $B_c$ production offers a promising way of shedding light
over these discrepancies and gaining an insight on the underlying physics.
In proton-proton ($pp$) collisions at the Large Hadron Collider (LHC),
$B_c$ mesons are expected to be mainly produced through the gluon-gluon fusion process
$gg\rightarrow B_c +b+\bar{c}$.
The production cross-sections
of the $B_c$ mesons have been calculated
in the fragmentation approach~\cite{Braaten:1993jn,Cheung:1999ir}
and in the complete
order-$\alpha_s^4$ approach~\cite{Chang:1992jb,Chang:1994aw,Chang:1996jt,Chang:2005bf,Kolodziej:1995nv,Berezhnoy:1994ba,Berezhnoy:1996an,Baranov:1997wy},
where $\alpha_s$ is the strong-interaction coupling.
In the latter approach,
the total production cross-section of the $B_c$ ground state, $\Bc$,
at a center-of-mass energy of 8\tev, integrated over the whole phase space and including
contributions from intermediate excited states,
is predicted to be about 0.2\%~\cite{Chang:2003cr,Gao:2010zzc} of the inclusive $b\bar{b}$ cross-section~\cite{LHCb-PAPER-2011-003}.

Previously, only the average ratios of $\Bc$ to $\Bu$ or $\Bs$
cross-sections in specific kinematic regions
had been
measured~\cite{Abe:1998wi,*Abe:1998fb,LHCB-PAPER-2012-028,LHCb-PAPER-2013-044},
and double-differential cross-sections have not yet been measured.
The production cross-sections of $b$-hadrons show different
transverse momentum dependencies~\cite{Aaltonen:2008eu,LHCb-PAPER-2011-018,LHCB-PAPER-2014-004,LHCB-PAPER-2014-002}.
A precise measurement of $\Bc$ production
as a function of transverse momentum and rapidity
will provide useful information on the largely unknown
production mechanism of the \Bc\ meson and
other bound states of heavy quarks,
and is also important to guide \Bc\ studies at the LHC.

In this Letter we report on the first measurement
of the ratio of double differential inclusive
production cross-sections multiplied by branching fractions,
\begin{eqnarray}
R(\ptrans,y)\equiv \frac{{\rm d}\sigma_{\Bc}(\ptrans,y)\;{\cal B}(\bcjpsipi)}{{\rm d}
\sigma_{\bplus}(\ptrans,y)\;{\cal B}(\bpjpsik)},
\label{eq:Rdef}
\end{eqnarray}
where transverse momentum $\pt$ and rapidity $y$ refer to the $b$ meson.
The cross-section includes contributions from excited states.
We use a sample of $pp$ collision
data at 8\tev, corresponding to an integrated luminosity
of 2.0\invfb recorded by the LHCb experiment.
The \Bc\ and \bplus\ mesons are reconstructed in the exclusive decays
\bcjpsipi\ and \bpjpsik\ respectively,  with \jpsimumu.
The inclusion of charge conjugate modes is implied throughout this Letter.

The \lhcb detector~\cite{Alves:2008zz} is a single-arm forward
spectrometer covering the \mbox{pseudorapidity} range $2<\eta <5$,
designed for the study of particles containing \bquark or \cquark
quarks. The detector includes a high-precision tracking system
consisting of a silicon-strip vertex detector surrounding the $pp$
interaction region,
a large-area silicon-strip detector located
upstream of a dipole magnet with a bending power of about
$4{\rm\,Tm}$, and three stations of silicon-strip detectors and straw
drift tubes
placed downstream of the magnet.
The combined tracking system provides a momentum measurement with
a relative uncertainty that varies from 0.4\% at low momentum, \ptot, to 0.6\% at 100\gevc.
The minimum distance of a track to a primary vertex, the impact parameter (IP),
is measured with a resolution of $(15+29/\pt)\mum$,
where \pt is in \gevc.
Different types of charged hadrons are distinguished using information
from two ring-imaging Cherenkov detectors.
Photon, electron and
hadron candidates are identified by a calorimeter system consisting of
scintillating-pad and preshower detectors, an electromagnetic
calorimeter and a hadronic calorimeter. Muons are identified by a
system composed of alternating layers of iron and multiwire
proportional chambers.

The trigger consists of a
hardware stage, based on information from the calorimeter and muon
systems, followed by a software stage, in which all charged particles
with $\ptrans>300\mevc$ are reconstructed~\cite{LHCb-DP-2012-004}.
Events
are first required to pass the
hardware trigger, which requires one or two
muons with high $\pt$. In the subsequent software trigger,
the event is required to have one muon with
high $\pt$ and large IP with respect to all
primary $pp$ interaction vertices~(PVs),
or a pair of oppositely charged muons with an invariant
mass consistent with the known \jpsi\ meson mass~\cite{PDG2014}.
Finally, the tracks of two or more of the final state
particles are required to form a vertex that is significantly
displaced from the PVs.
A multivariate algorithm~\cite{BBDT} is also used to
identify secondary vertices consistent with the decay
of a $b$ meson.

The $b$-meson candidate selection
is performed in two steps, a preselection
and a final selection on the output of a multivariate
classifier based on a boosted decision tree
algorithm~(BDT)~\cite{Breiman,AdaBoost}.
Simulated \Bc\ and \Bu\ decays are used to optimize
the $b$-meson candidate selection.
Production of $\Bu$ mesons is simulated using
\pythia~6.4~\cite{Sjostrand:2006za} with an
LHCb specific configuration~\cite{LHCb-PROC-2010-056}.
The generator \bcvegpy~\cite{Chang:2005hq}
is used to simulate $\Bc$ meson production.
Decays of $\Bc$, $\Bu$ and $\jpsi$ mesons
are described by \evtgen~\cite{Lange:2001uf}
and photon radiation is simulated using
the \photos\ package~\cite{Golonka:2005pn}.
The decay products are traced through the detector by the \geant
package~\cite{Allison:2006ve,*Agostinelli:2002hh,LHCb-PROC-2011-006}.
Following Ref.~\cite{LHCb-PAPER-2013-063},
the \Bc\ meson lifetime is set to
$\tau_{\Bc} = 0.509 $~ps.
The selection requirements are
the same for $\bcjpsipi$ and $\bpjpsik$ candidates.

In the preselection,
\jpsi\ candidates are formed from pairs of oppositely charged particles
with \ptrans\ larger than
0.55\gevc, with a good quality of the track fit
and identified as muons.
The two muons are required to originate from a common vertex.
The \jpsi\ candidates with invariant mass
between $3.04\gevcc$ and $3.14\gevcc$ are combined
with a charged particle
that has $\pt>1.0\gevc$, a good quality of the track fit and
is separated from any PV.
The pion mass hypothesis is assigned to the track for the
selection of the \Bc\ candidate and the kaon hypothesis for
that of the \bplus\ candidate.
The \jpsi\ candidate and the hadron ($\pi$ or $K$) are required
to originate from a common vertex.
To improve the $b$-meson mass resolution,
the mass of the muon pair is constrained to the
known \jpsi\ meson mass~\cite{PDG2014} in this vertex fit.
The $b$-meson candidates are required to have
a decay time larger than $0.2\,{\rm ps}$,
and to point toward the primary vertex.

In the final selection,
the BDT is trained
using a simulated $\Bc$ signal sample and background events populating the
data mass sideband $6376<M_{\jpsi\pi^+}<6600\,\mevcc$.
The following variables are used as input to the BDT:
$\chi^2_{\rm IP}$ of all particles;
$\pt$ of muons, $J/\psi$ and $\pi^+$;
and the $b$-meson decay length, decay time, and the vertex fit
$\chi^2$ of a fit to the decay tree~\cite{Hulsbergen:2005pu}.
The quantity $\chi^2_{\rm IP}$ is defined as the difference in
$\chi^2$ of a given primary vertex reconstructed with and without the
considered particle.
The selection value on the BDT output is chosen to maximize the signal
significance $N_{\rm S}/\sqrt{N_{\rm S}+N_{\rm B}}$,
where $N_{\rm S}$ and $N_{\rm B}$ are the expected numbers of
signal and background events, respectively.
The same BDT requirements are used for the $\Bu$ meson.

The \Bc\ and \bplus\ candidates are subdivided into 10 bins of \ptrans\ and 3 bins of $y$.
Bin sizes are chosen to contain approximately the same number of
signal candidates,
except for the highest $\pt$ bin.
The differential production ratio $R$ is measured as
\begin{eqnarray}
R(\ptrans,y)=\frac{N_{\Bc}(\ptrans,y)}{N_{\bplus}(\ptrans,y)} \frac{\epsilon_{\bplus}(\ptrans,y)}{\epsilon_{\Bc}(\ptrans,y)},
\label{eq:diffxsEq}
\end{eqnarray}
where $N_{B}(\ptrans,y)$ is the number of reconstructed signal decays and
$\epsilon_{B}(\ptrans,y)$
is the total efficiency in a given (\ptrans,$y$) bin,
including geometrical acceptance, reconstruction, selection and trigger
effects.

In each $\ptrans$ and $y$ bin,
the number of signal decays is determined by performing
an extended maximum likelihood fit to the unbinned invariant
mass distribution of $\Bc$ candidates reconstructed in
$6150<M_{\jpsi\pi^+}<6550\mevcc$
and $\Bu$ candidates in
$5150<M_{\jpsi K^+}<5550\mevcc$.
For both $\Bc \to \jpsi \pip$ and $\Bu \to \jpsi K^+$ decays,
the fit includes components for
signal, combinatorial background,
and Cabibbo-suppressed backgrounds \bcjpsik\ and $\bpjpsipi$.
Other sources of backgrounds, such as $\Bc\to\jpsi\mup\nu_{\mu}$, are
negligible.
The \bcjpsipi\ signal
is described by a
double-sided Crystal Ball (DSCB) function,
which is an empirical function with a Gaussian core and power-law tails on both sides.
The \bpjpsik\ signal
is described by the sum of two DSCB functions,
to account for different mass resolutions in different kinematic
regions.
The tail parameters are determined from simulation.
The combinatorial background is described by an exponential function.
The shapes of the Cabibbo-suppressed backgrounds
are determined from simulation. The ratios of the yield
of the Cabibbo-suppressed background to that of the signal are fixed
to the central value of
$\mathcal{B}(\bcjpsik)/\mathcal{B}(\bcjpsipi)=
(6.9 \pm 2.0)\%$ for $\Bc$ candidates~\cite{LHCB-PAPER-2013-021}, and
$\mathcal{B}(B^+\to \jpsi \pi^+)/\mathcal{B}(B^+\to \jpsi
K^+)=(3.83 \pm 0.13)\%$ for $\Bu$ candidates~\cite{LHCB-PAPER-2011-024},
respectively.

\begin{figure*}
  \centering
  \includegraphics[width=0.45\textwidth]{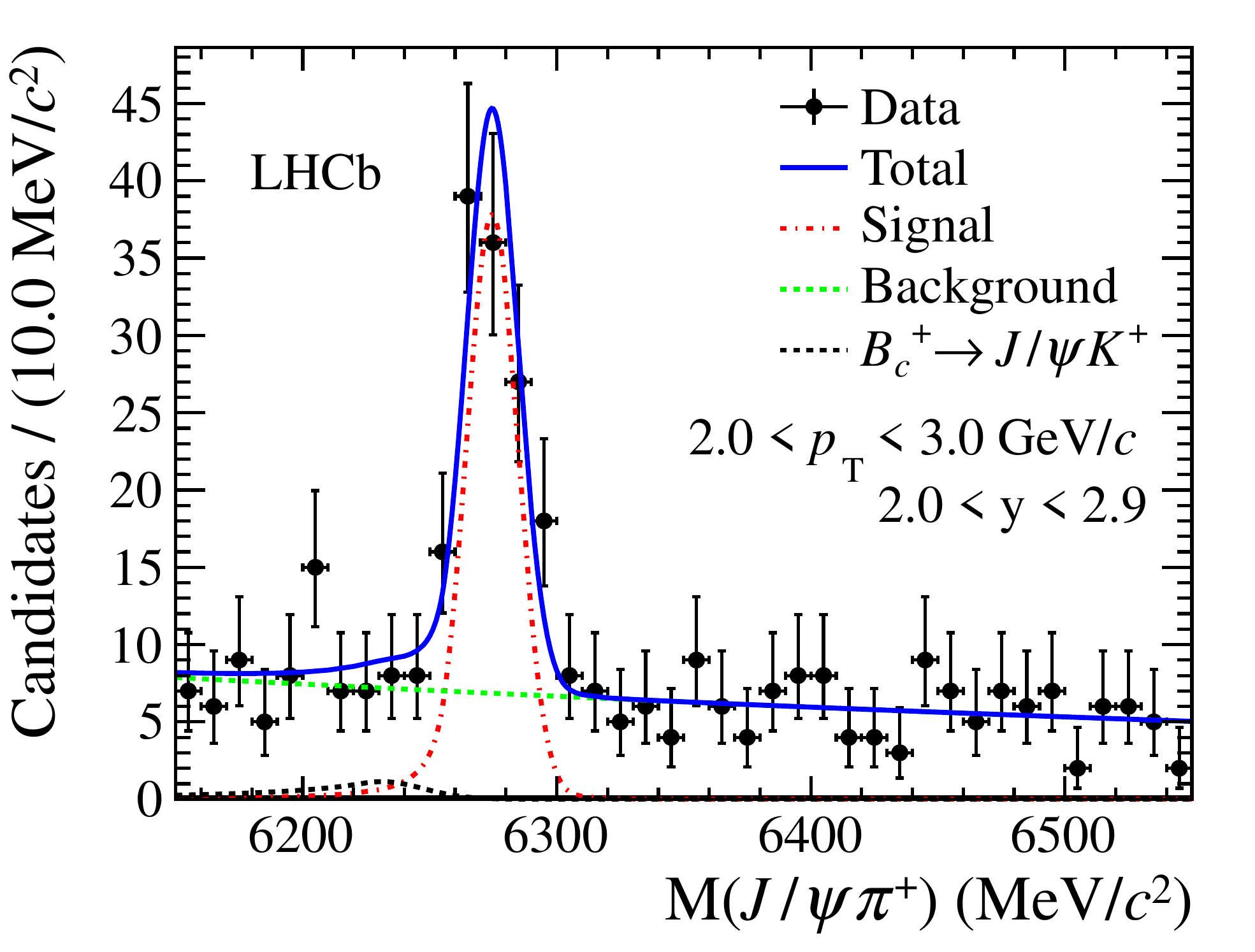}
  \includegraphics[width=0.45\textwidth]{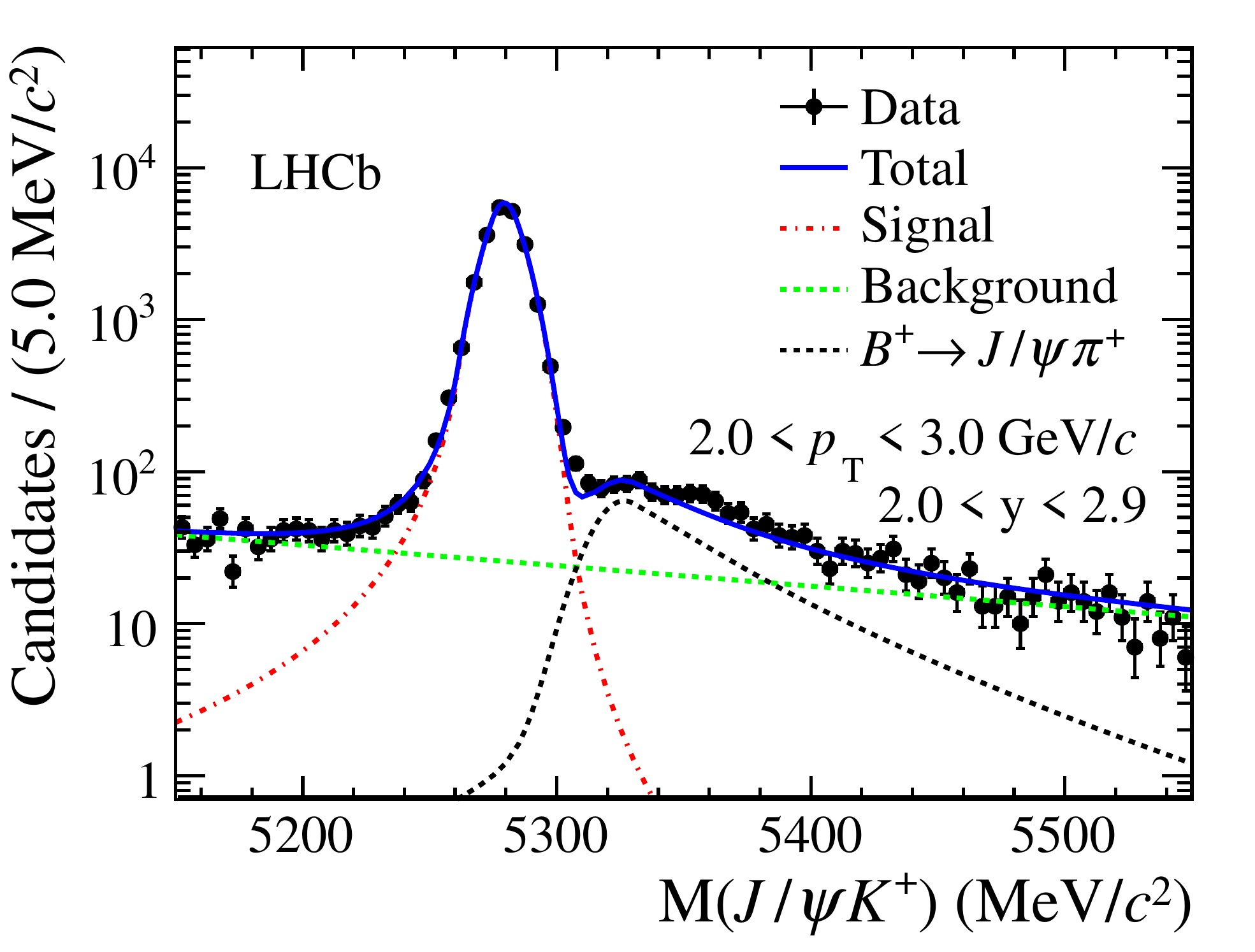}
  \caption{Invariant mass distribution of ({\it left}) $\bcjpsipi$ and ({\it right})
$\Bu\to\jpsi K^+$ candidates with $2.0<\pt<3.0\gevc$ and $2.0<y<2.9$. The results
of the fit described in the text are superimposed.}
  \label{fig:Bmasses}
%\end{figure*}
%
%\begin{figure*}[!h]
  \centering
  \includegraphics[width=0.45\textwidth]{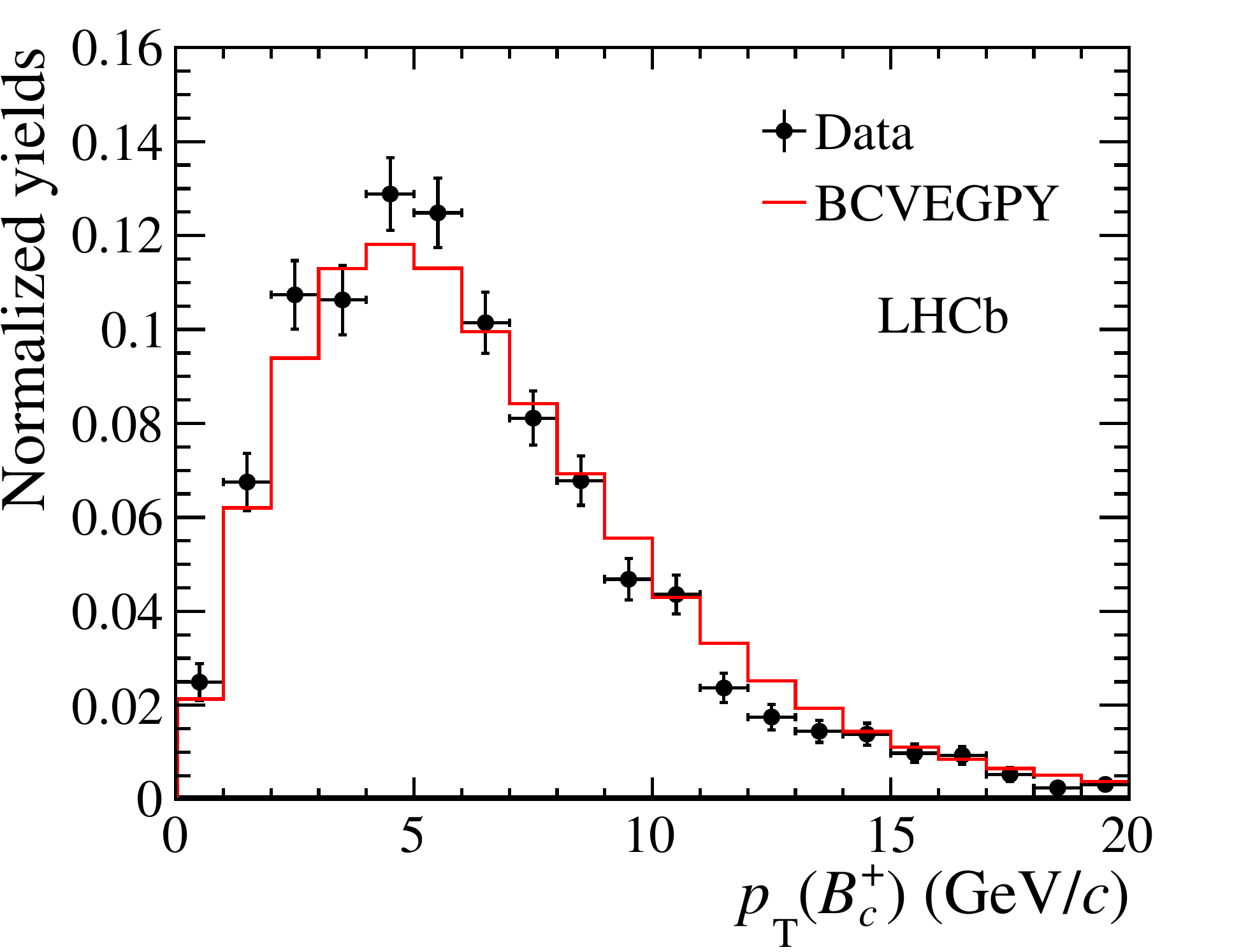}
  \includegraphics[width=0.45\textwidth]{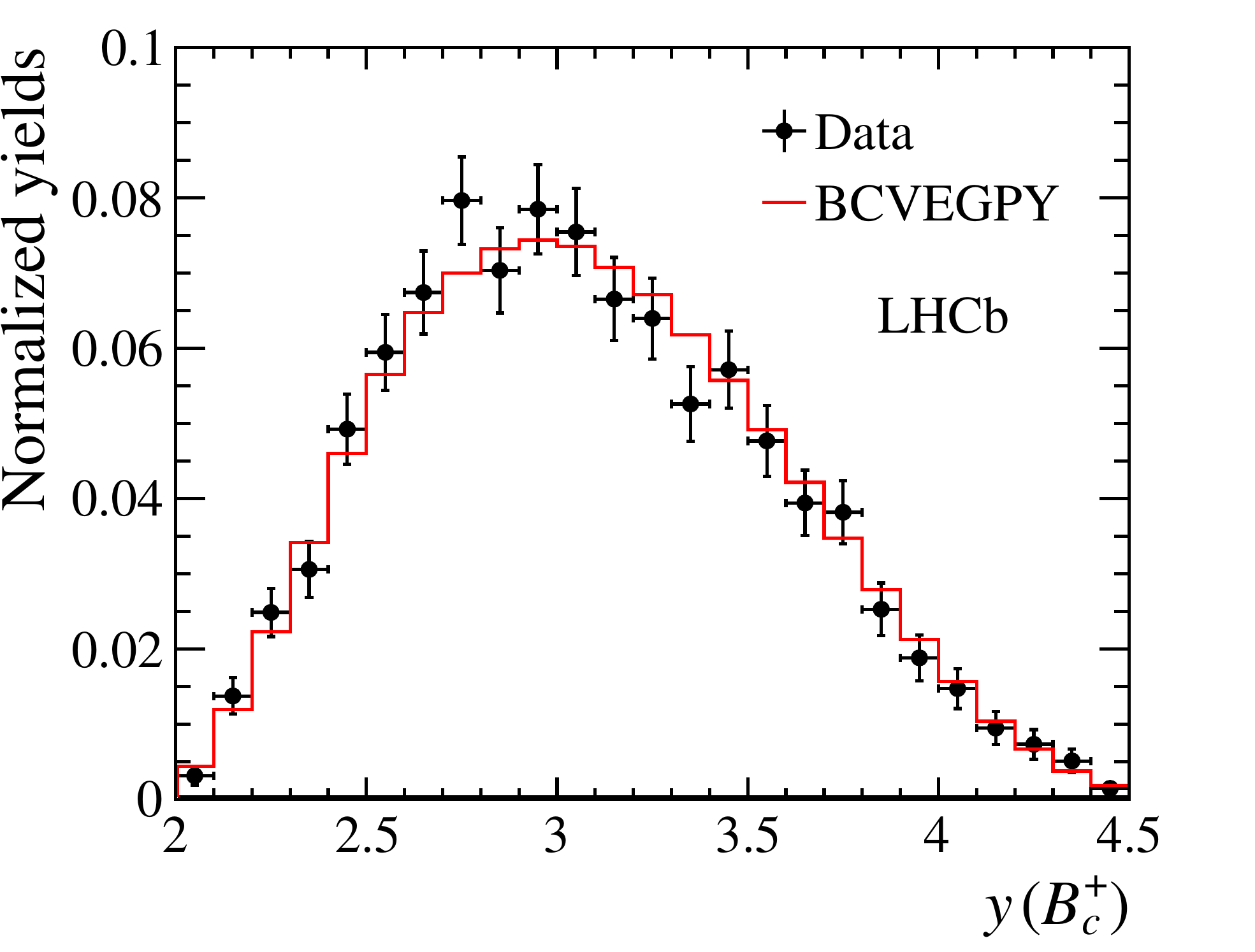}
  \caption{Distributions of ({\it left}) $\pt$ and ({\it right}) $y$ of
    the $\Bc$ signal after event selection.
    The points with error bars are background-subtracted data, and the
    solid histogram is the simulation based on the complete
    order-$\alpha_s^4$ calculation, implemented in the $\Bc$ generator
    \bcvegpy~\cite{Chang:2005hq}.
    The uncertainties are statistical.  }
  \label{fig:PTYComp}
\end{figure*}

As an example, Fig.~\ref{fig:Bmasses} shows
the $\Bc$ and $\Bu$ mass distributions
together with the fit results for the
bin $2.0<\pt<3.0\gevc$ and $2.0<y<2.9$.
The mass resolution is approximately 11\mevcc for $\Bc$ signals
and 8.7\mevcc for $\Bu$ signals.
Summing over all bins,
a total signal yield of $3.1\times 10^3$ $\Bc$ candidates
and $7.1 \times 10^5$ $\Bu$ candidates
is obtained.
In each (\ptrans,$y$) bin the total efficiency is determined from
simulation and ranges from 2.4\% to 23.2\% for \Bc\ candidates and
from 3.6\% to 33.5\% for \bplus\ candidates.

The systematic uncertainties associated with the signal
shape in each bin (0.1\% -- 2.6\%) are estimated by
comparing the ratios between input signal yields and fit results in
simulation.
The uncertainties from
the combinatorial background shape (0.1\% -- 4.4\%) are determined by
varying the fit function.
The input value for the ratio of branching fractions
$\mathcal{B}(\bcjpsik)/\mathcal{B}(\bcjpsipi)$ is
varied within its uncertainty and
the resulting difference (0.1\% -- 0.9\%) is taken as systematic uncertainty.
The effect of $\mathcal{B}(\bpjpsipi)/\mathcal{B}(\bpjpsik)$ is found to be negligible.
The systematic uncertainty associated with the relative trigger efficiency
is estimated to be 1\%.
Other effects, such as the $(\pt, y)$ binning scheme,
the shapes of the Cabibbo-suppressed backgrounds,
the $\Bc$ lifetime uncertainty and
the uncertainty of tracking efficiency,
are negligible.

\begin{figure}
 \centering
  \includegraphics[width=0.49\textwidth]{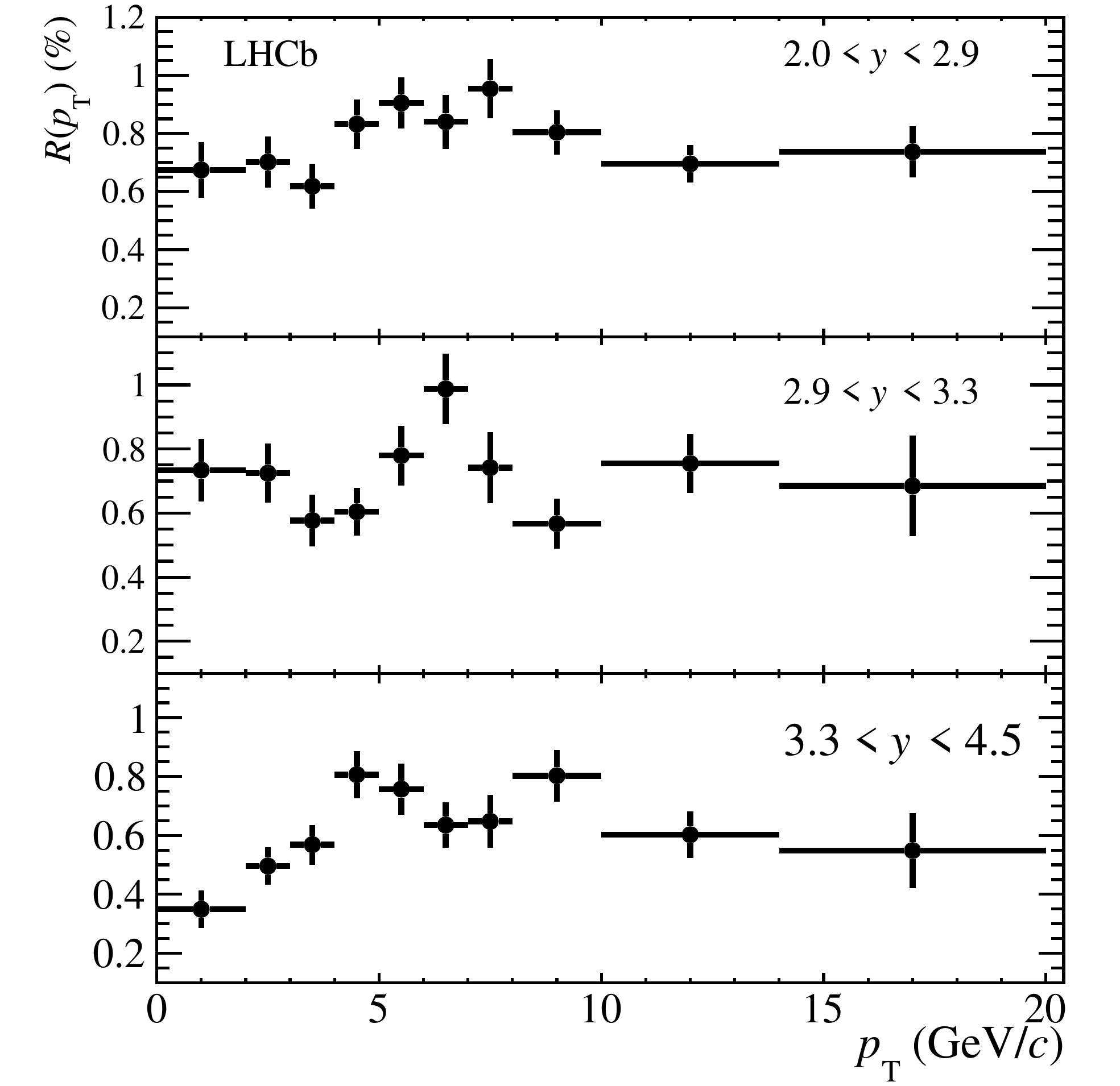}
  \caption{Ratio $R(\ptrans,y)$ as a function of $\pt$ in the regions
   ({\it top}) $2.0<y<2.9$,
   ({\it middle}) $2.9<y<3.3$, and
   ({\it bottom}) $3.3<y<4.5$.
   The error bars on the data show the statistical and systematic uncertainties added in quadrature. }
  \label{fig:BinResults}
\end{figure}

\begin{figure*}
  \includegraphics[width=0.45\textwidth]{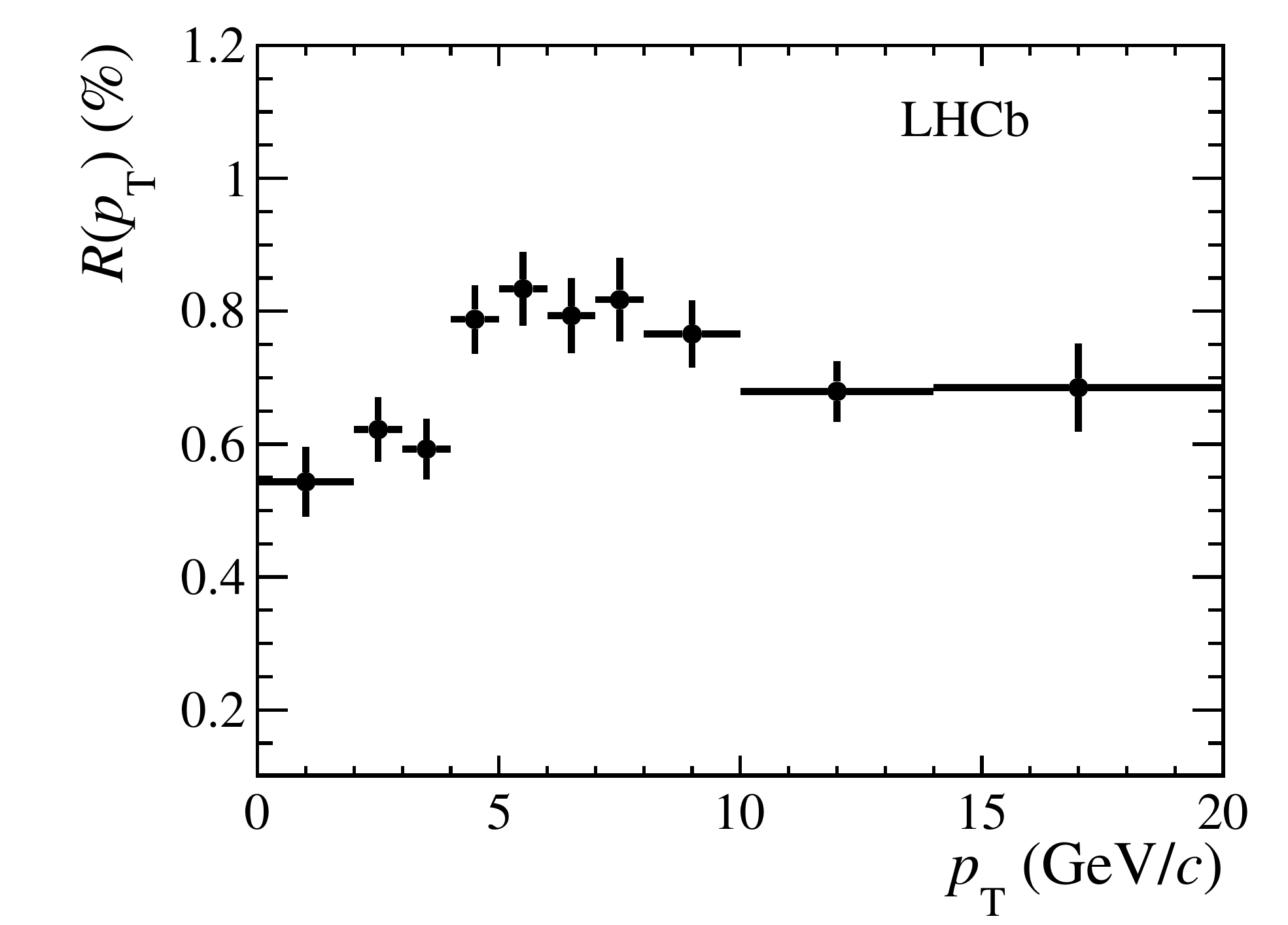}
  \includegraphics[width=0.45\textwidth]{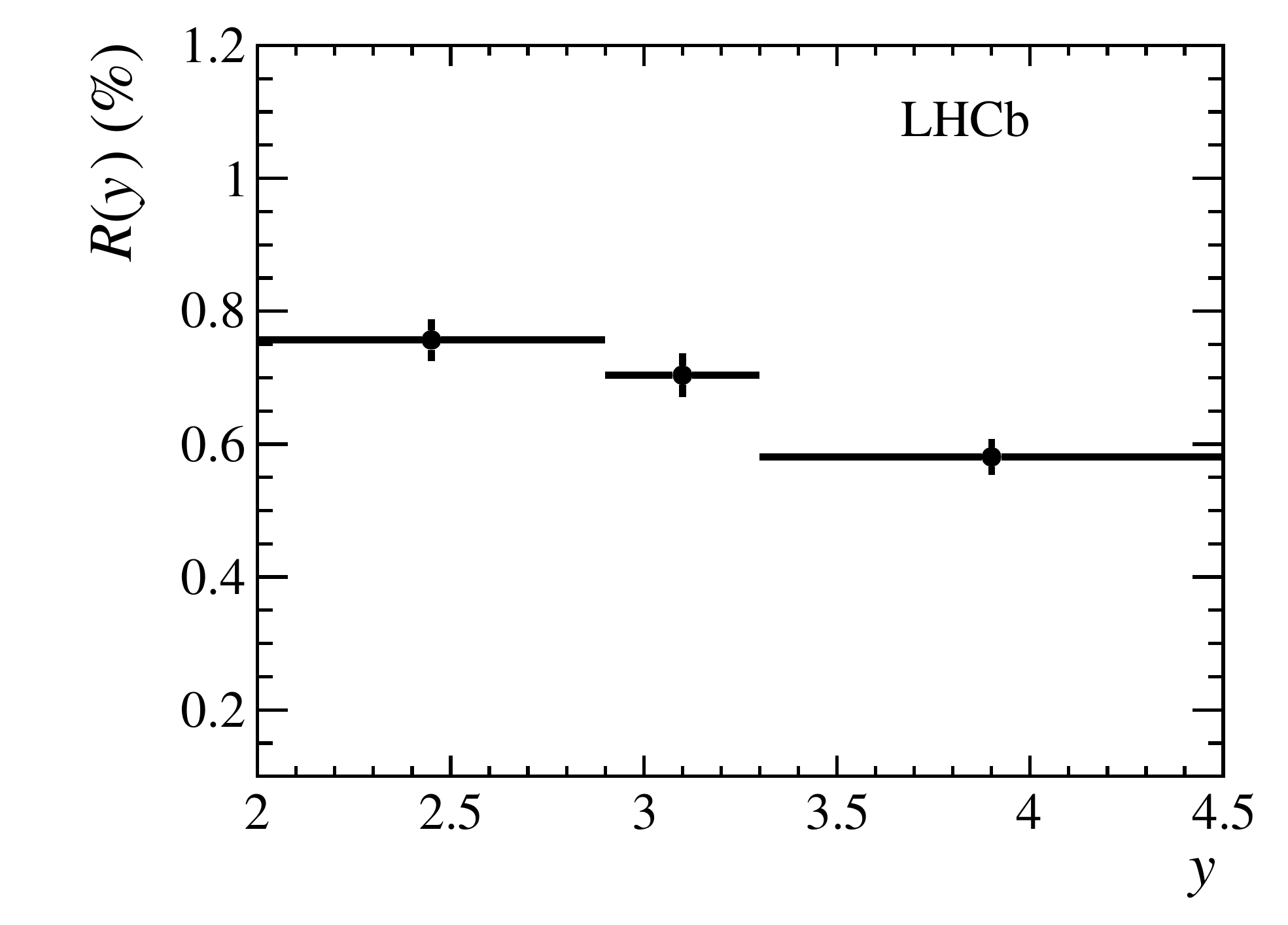}
  \caption{Ratio ({\it left}) $R(\ptrans)$ as a function of $\ptrans$
    integrated over $y$ in the region $2.0<y<$4.5 and ({\it right})
    $R(y)$ as a function of $y$ integrated over $\ptrans$ in the
    region $0<\pt<20\gevc$.
    The error bars on the data show the statistical and systematic uncertainties added in quadrature.  }
  \label{fig:Results}
\end{figure*}

Figure~\ref{fig:PTYComp} shows that simulation
provides a good description of
$\pt$ and $y$ distributions of $\Bc$ mesons in data.
The values of $R(\ptrans,y)$ in the range
$0<\pt<20$\gevc and $2.0<y<4.5$ are shown in Fig.~\ref{fig:BinResults}
and Ref.~\cite{suppmat}.
Figure~\ref{fig:Results} shows the ratio
$R(\ptrans)$ integrated over $y$ in the region $2.0<y<4.5$
and $R(y)$ integrated over $\ptrans$ in the
region $0<\pt<20\gevc$.
The ratios are found to vary as a function of $\pt$ and $y$.
The results are compared with the theoretical predictions in Ref.~\cite{suppmat}.

The resulting integrated value of $R$ in the region
 $0<\ptrans<20\gevc$ and $2.0<y<4.5$ is measured to be
$$R=\ratiotot,$$
where the first uncertainty
is statistical and the second systematic.
To enable comparison with the previous LHCb
measurement~\cite{LHCB-PAPER-2012-028},
$R$ and its total uncertainty are also reported in the range
$4<\ptrans<20\gevc$ and $2.5<\eta<4.5$
as ($0.698\pm0.023$)\%.
The previous LHCb measurement of $R$ at 7 TeV of
Ref.~\cite{LHCB-PAPER-2012-028}
is updated using the recent measurement of the $\Bc$ lifetime~\cite{LHCb-PAPER-2013-063}
to be $(0.61\pm0.12)\%$.

In summary, we present the first measurement of the $\Bc$ double differential
production cross-section ratio with respect to that of the $\Bu$
meson. The measurement is performed
in three bins of rapidity and ten bins of $\ptrans$ in $pp$
collisions at $\sqrt{s}=8$\tev on a data sample collected with the
LHCb detector.
The relative production rates of $\Bc$ and $\Bu$ mesons are found to
depend on their transverse momentum and rapidity.
The measured transverse momentum and rapidity distributions of
the \Bc\ meson are well described by the complete order-$\alpha_s^4$ calculation.
However, the theoretical predictions on the \Bc\ and \bplus\ production cross-sections
suffer from big uncertainties~\cite{Chang:2003cr,Cacciari:2012ny},
and the prediction of the branching fraction of
the \bcjpsipi\ decay has a big spread (see for example Ref.~\cite{Qiao:2012hp}), more work on the theoretical
side is required to have concluding remarks on the \Bc\ absolute production rate.
These results will provide useful information on the \Bc\ production mechanism,
and help understand the quarkonium production, therefore deepen our understanding of QCD.
\\
\\
\noindent
We express our gratitude to our colleagues in the CERN
accelerator departments for the excellent performance of the LHC. We
thank the technical and administrative staff at the LHCb
institutes. We acknowledge support from CERN and from the national
agencies: CAPES, CNPq, FAPERJ and FINEP (Brazil); NSFC (China);
CNRS/IN2P3 (France); BMBF, DFG, HGF and MPG (Germany); SFI (Ireland); INFN (Italy);
FOM and NWO (The Netherlands); MNiSW and NCN (Poland); MEN/IFA (Romania);
MinES and FANO (Russia); MinECo (Spain); SNSF and SER (Switzerland);
NASU (Ukraine); STFC (United Kingdom); NSF (USA).
The Tier1 computing centres are supported by IN2P3 (France), KIT and BMBF
(Germany), INFN (Italy), NWO and SURF (The Netherlands), PIC (Spain), GridPP
(United Kingdom).
We are indebted to the communities behind the multiple open
source software packages on which we depend. We are also thankful for the
computing resources and the access to software R\&D tools provided by Yandex LLC (Russia).
Individual groups or members have received support from
EPLANET, Marie Sk\l{}odowska-Curie Actions and ERC (European Union),
Conseil g\'{e}n\'{e}ral de Haute-Savoie, Labex ENIGMASS and OCEVU,
R\'{e}gion Auvergne (France), RFBR (Russia), XuntaGal and GENCAT (Spain), Royal Society and Royal
Commission for the Exhibition of 1851 (United Kingdom).
We would like to thank
Matteo Cacciari,
Chao-Hsi Chang,
Xing-Gang Wu,
and
Rui-Lin Zhu
for useful discussions.

\clearpage
\section*{Supplementary material}

Table~\ref{tab:bcyield} shows the ratio $R(\ptrans,y)$ in each $(\pt,y)$ bin.

\begin{table}[!hbp]
\tabcolsep 4mm
\begin{center}
\caption{\small \label{tab:bcyield}$R(\ptrans,y)$ in units of
  $10^{-2}$ as a function of \ptrans\ and $y$. The first uncertainty is statistical and the second systematic.}
\resizebox{\textwidth}{!}{%
\begin{tabular}{@{}r@{$\,\ptrans\,$}lccc|c@{}}
\toprule
\multicolumn{2}{c}{\ptrans(\gevc)}  & $2.0<y<2.9$ & $2.9<y<3.3$ & $3.3<y<4.5$ & $2.0<y<4.5$ \\
  \midrule
$ 0<$ & $<2 $ & $ 0.67\pm0.10\pm0.01 $ & $ 0.73\pm0.10\pm0.01 $ & $
0.35\pm0.06\pm0.01 $ & $ 0.54\pm0.05\pm0.01 $  \\
$ 2<$ & $<3 $ & $ 0.70\pm0.09\pm0.02 $ & $ 0.72\pm0.09\pm0.02 $ & $
0.50\pm0.06\pm0.01 $ & $ 0.62\pm0.05\pm0.01 $  \\
$ 3<$ & $<4 $ & $ 0.62\pm0.08\pm0.01 $ & $ 0.58\pm0.08\pm0.01 $ & $
0.57\pm0.07\pm0.02 $ & $ 0.59\pm0.05\pm0.01 $  \\
$ 4<$ & $<5 $ & $ 0.83\pm0.08\pm0.02 $ & $ 0.60\pm0.07\pm0.01 $ & $
0.81\pm0.08\pm0.02 $ & $ 0.79\pm0.05\pm0.01 $  \\
$ 5<$ & $<6 $ & $ 0.90\pm0.09\pm0.02 $ & $ 0.78\pm0.09\pm0.01 $ & $
0.76\pm0.09\pm0.02 $ & $ 0.83\pm0.06\pm0.01 $  \\
$ 6<$ & $<7 $ & $ 0.84\pm0.09\pm0.01 $ & $ 0.99\pm0.11\pm0.02 $ & $
0.64\pm0.08\pm0.01 $ & $ 0.79\pm0.06\pm0.01 $  \\
$ 7<$ & $<8 $ & $ 0.95\pm0.10\pm0.01 $ & $ 0.74\pm0.11\pm0.01 $ & $
0.65\pm0.09\pm0.01 $ & $ 0.82\pm0.06\pm0.01 $  \\
$ 8<$ & $<10 $ & $ 0.80\pm0.08\pm0.01 $ & $ 0.57\pm0.08\pm0.01 $ & $
0.80\pm0.09\pm0.02 $ & $ 0.77\pm0.05\pm0.01 $  \\
$ 10<$ & $<14 $ & $ 0.70\pm0.06\pm0.01 $ & $ 0.75\pm0.09\pm0.01 $ & $
0.60\pm0.08\pm0.01 $ & $ 0.68\pm0.05\pm0.01 $  \\
$ 14<$ & $<20 $ & $ 0.74\pm0.09\pm0.01 $ & $ 0.68\pm0.15\pm0.03 $ & $
0.55\pm0.13\pm0.02 $ & $ 0.68\pm0.07\pm0.01 $  \\
\midrule
$ 0<$ & $<20 $ & $ 0.76\pm0.03\pm0.01 $ & $ 0.70\pm0.03\pm0.01 $ & $
0.58\pm0.03\pm0.01 $ & $ 0.68\pm0.02\pm0.01 $  \\
\bottomrule
\end{tabular}
}
\end{center}
\end{table}

\begin{figure}[!htbp]
  % \centering
  \includegraphics[width=7.5cm]{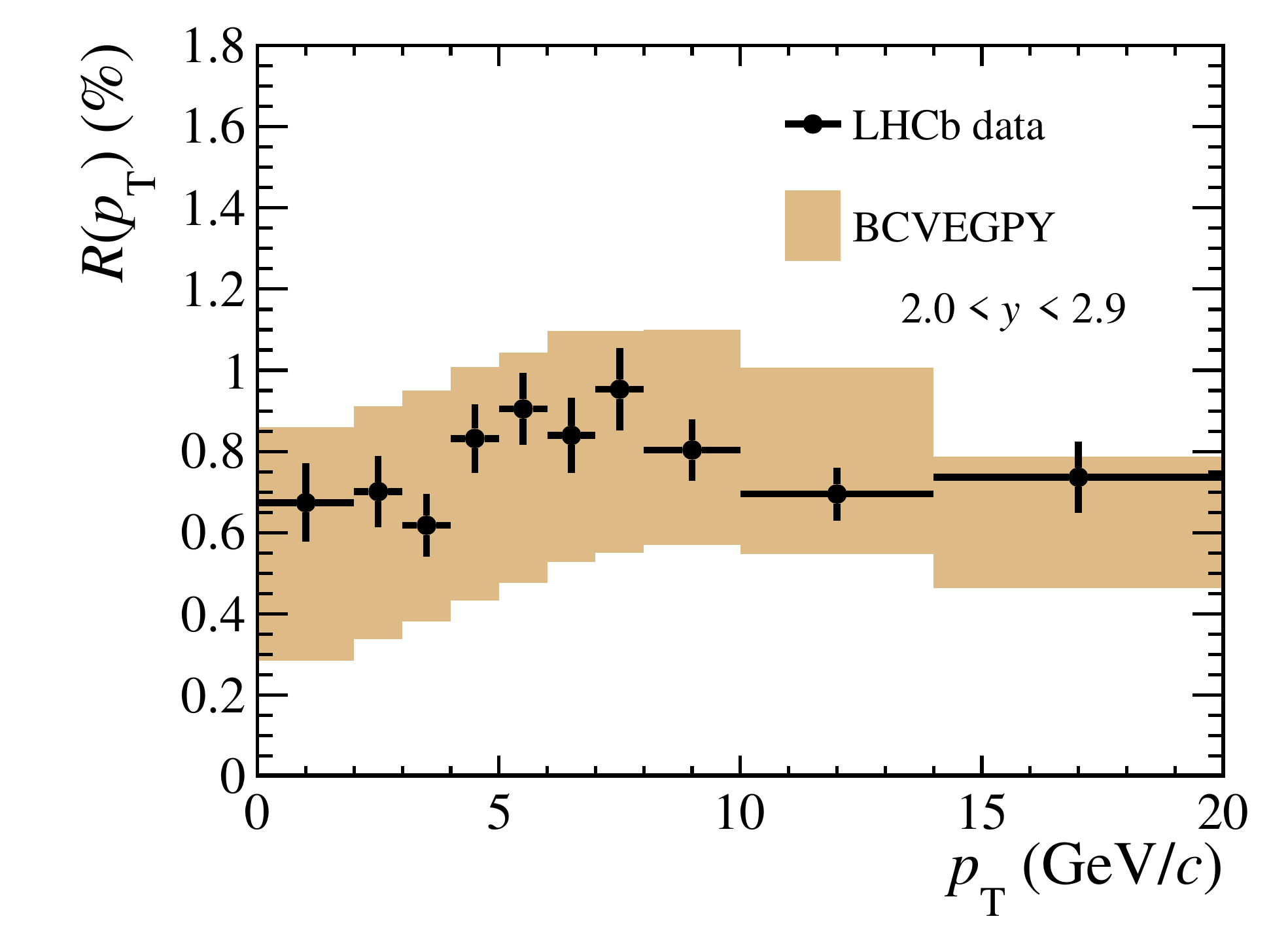}
  \includegraphics[width=7.5cm]{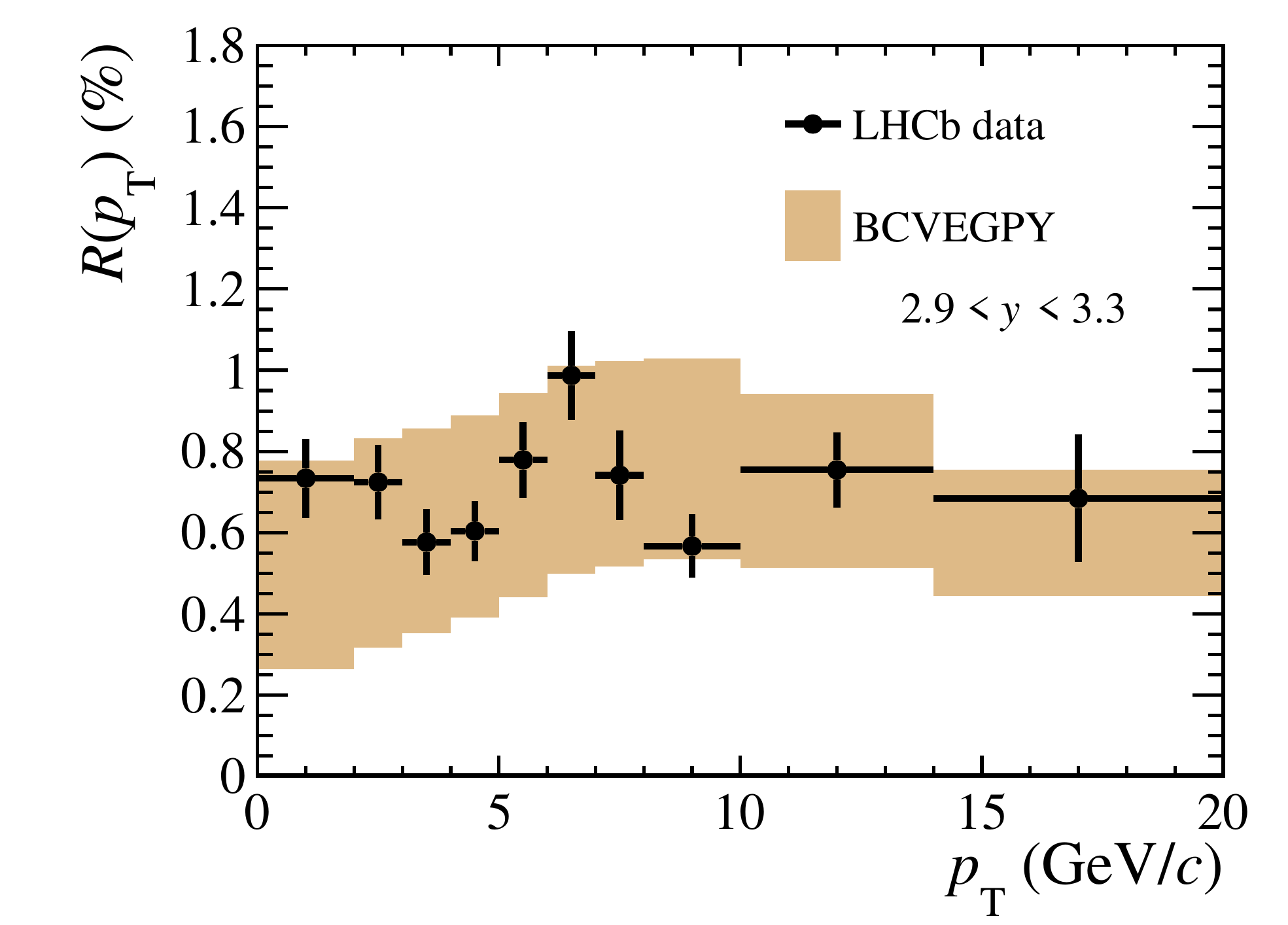}
  \includegraphics[width=7.5cm]{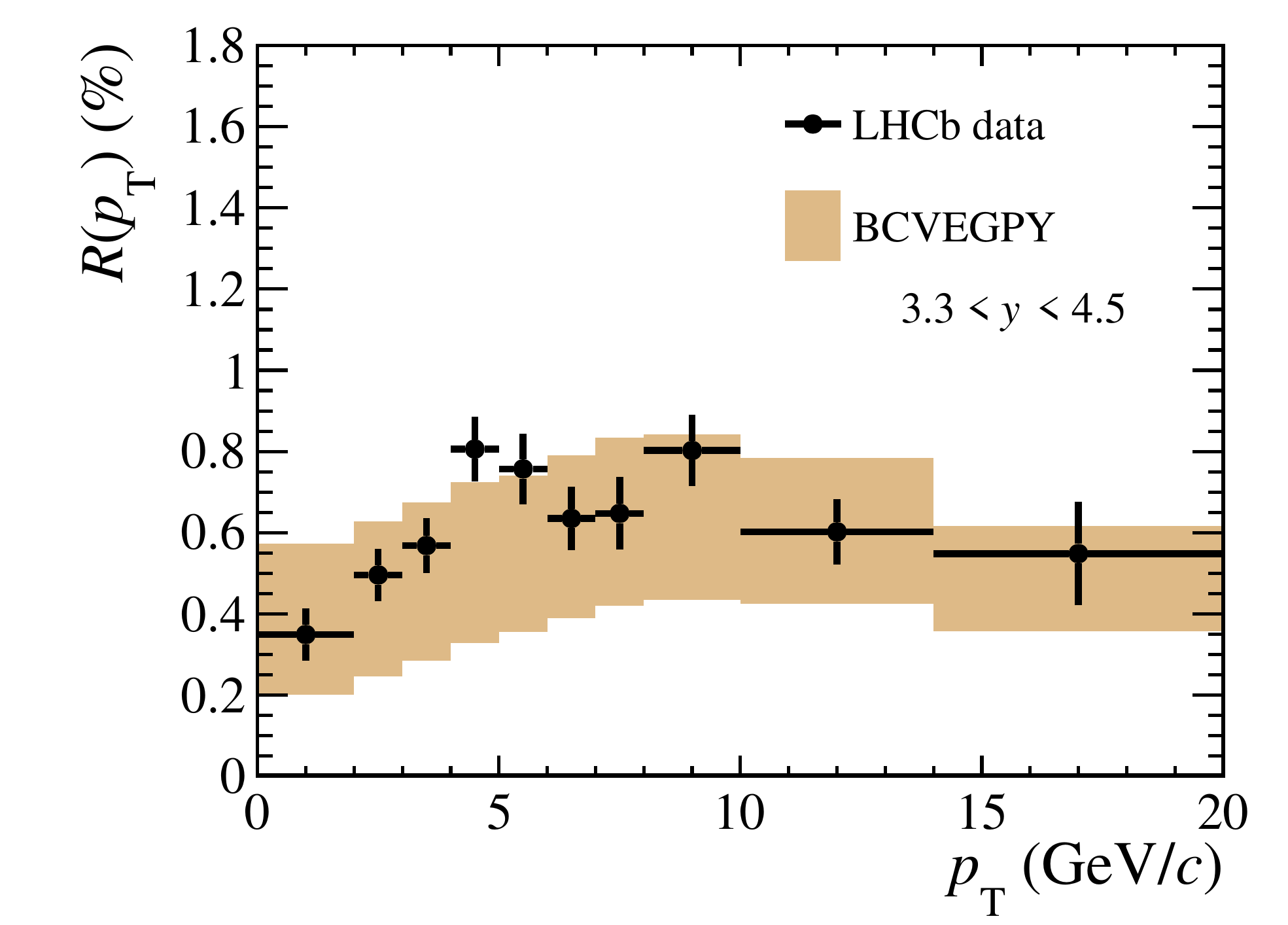}
  \caption{Ratio $R(\ptrans,y)$ as a function of $\pt$ in the regions
    $2.0<y<2.9$ ({\it top left}),
    $2.9<y<3.3$ ({\it top right}), and
    $3.3<y<4.5$ ({\it bottom left}),
    with theoretical predictions
    following the $\alpha_s^4$ approach~\cite{Chang:2005hq} overlaid.}
  \label{fig:res_theorycomp}
\end{figure}

\begin{figure}[!htbp]
  \centering
  \includegraphics[width=7.5cm]{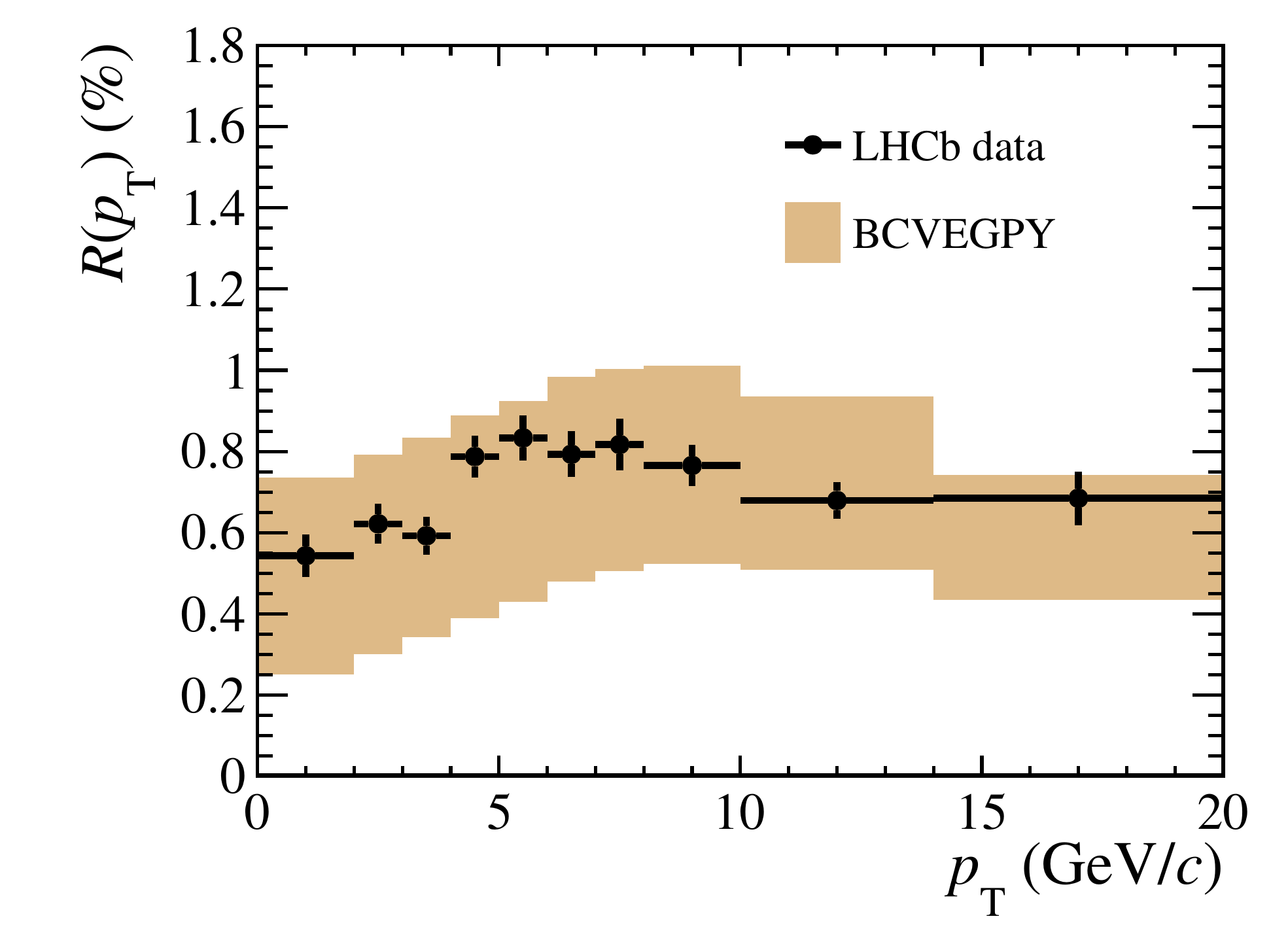}
  \includegraphics[width=7.5cm]{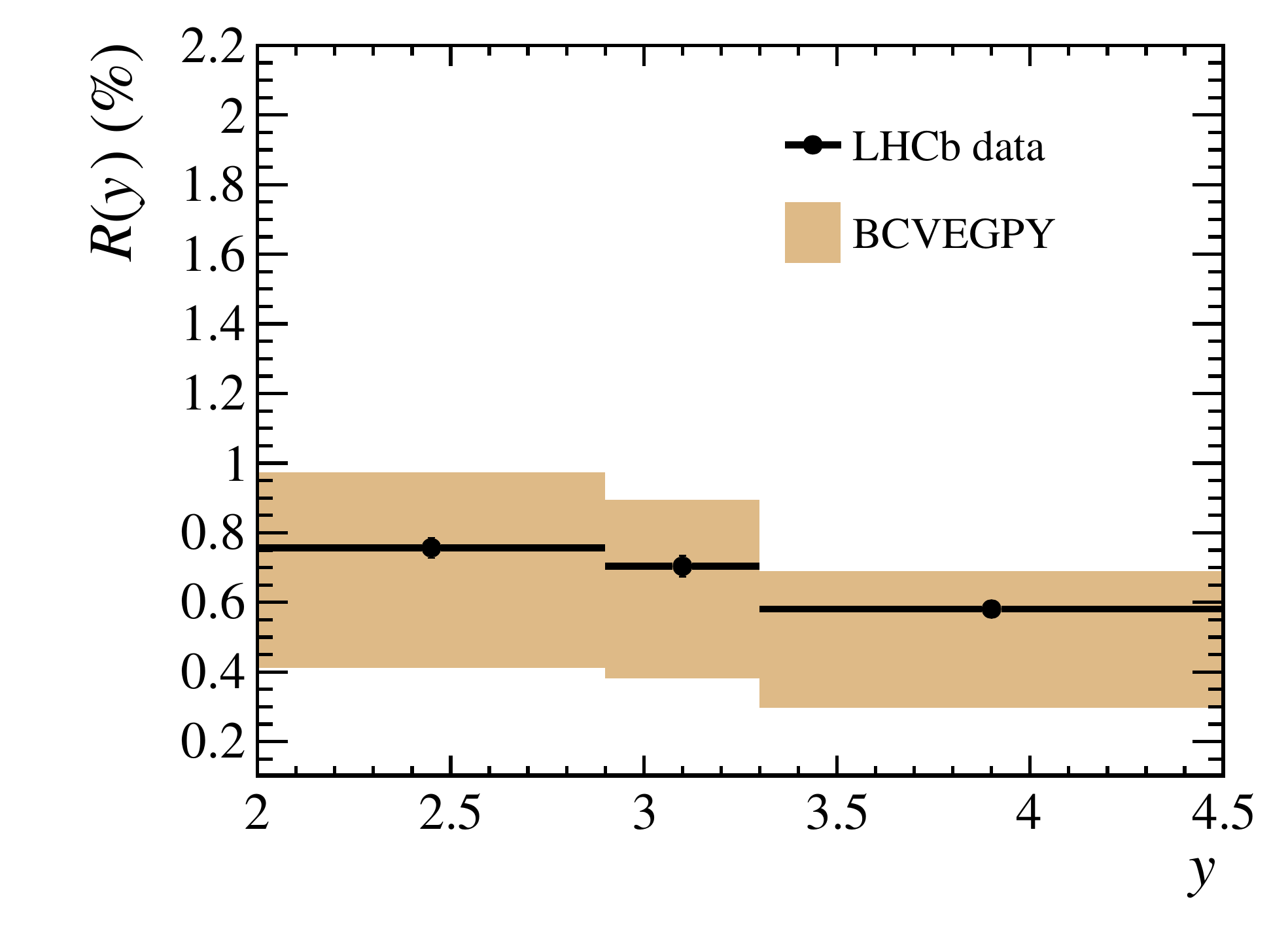}
  \caption{Ratio $R(\ptrans)$ as a function of $\ptrans$ integrated over $y$ in the region 2.0$<y<$4.5
    ({\it left}) and $R(y)$ as a function of $y$ integrated over $\ptrans$ in the region 0$<\pt<20\gevc$
    ({\it right}) are compared to the theoretical predictions
    following the $\alpha_s^4$ approach~\cite{Chang:2005hq}.
  }
  \label{fig:pt_y_rescomp}
\end{figure}

The results are compared with the
theoretical predictions in
Fig.~\ref{fig:res_theorycomp} and Fig.~\ref{fig:pt_y_rescomp}.
For $\Bc$ meson the predictions
following the
$\alpha_s^4$ approach~\cite{Chang:2005hq} %~\cite{Chang2006241}%{Chang:2005hq}
are shown.
We use the CTEQ6LL~\cite{Pumplin:2002vw} parton distribution functions,
and the leading order running $\alpha_s$, the characteristic energy
scale $Q^2 = \pt^2 + m_{\Bc}^2$, and the masses of the $b$ and $c$
quarks are set to $m_b=4.95\gevcc$ and $m_c=1.326\gevcc$.
The normalization of the theoretical predictions uses
$0.47\mub$ as the \Bc\ production cross-section in the whole phase space
and 0.33\% for $\BR(\bcjpsipi)$~\cite{Qiao:2012hp},
corrected for the latest measurement of the $\Bc$ lifetime.
% and another following the fragmentation approach~\cite{Berezhnoy:2013cda}.
The  theoretical prediction on the $\Bu$ cross-section is based on the
fixed order + next-to-leading log (FONLL) framework~\cite{Cacciari:2012ny}.
The uncertainties on the theory curves are the
uncertainties of the FONLL calculation,
including the uncertainties of the $b$ quark mass,
the renormalisation and factorisation scales, and
CTEQ6.6~\cite{Nadolsky:2008zw} functions.
The FONLL predictions are scaled according to
the measured branching fraction value $\BR(\Bu\to\jpsi K^+)=0.106\%$~\cite{PDG2014}
and the \bplus\ production cross-section $38.9\mub$
measured at $\sqrt{s}= 7\tev$~\cite{LHCb-PAPER-2013-004}
increased by 20\% due to higher collision energy~\cite{LHCb-PAPER-2013-016}.

%%%%%%
%%%%%%

\clearpage

\addcontentsline{toc}{section}{References}
\setboolean{inbibliography}{true}
\bibliographystyle{LHCb}
\bibliography{main,BcRev,LHCb-PAPER,LHCb-CONF,LHCb-DP,LHCb-TDR}

\ifx\mcitethebibliography\mciteundefinedmacro
\PackageError{LHCb.bst}{mciteplus.sty has not been loaded}
{This bibstyle requires the use of the mciteplus package.}\fi
\providecommand{\href}[2]{#2}
\begin{mcitethebibliography}{10}
\mciteSetBstSublistMode{n}
\mciteSetBstMaxWidthForm{subitem}{\alph{mcitesubitemcount})}
\mciteSetBstSublistLabelBeginEnd{\mcitemaxwidthsubitemform\space}
{\relax}{\relax}

\bibitem{Brambilla:2010cs}
N.~Brambilla {\em et~al.}, \ifthenelse{\boolean{articletitles}}{{\it {Heavy
  quarkonium: progress, puzzles, and opportunities}},
  }{}\href{http://dx.doi.org/10.1140/epjc/s10052-010-1534-9}{Eur.\ Phys.\ J.\
  {\bf C71} (2011) 1534}, \href{http://arxiv.org/abs/1010.5827}{{\tt
  arXiv:1010.5827}}\relax
\mciteBstWouldAddEndPuncttrue
\mciteSetBstMidEndSepPunct{\mcitedefaultmidpunct}
{\mcitedefaultendpunct}{\mcitedefaultseppunct}\relax
\EndOfBibitem
\bibitem{Chang:1979nn}
C.-H. Chang, \ifthenelse{\boolean{articletitles}}{{\it {Hadronic production of
  $J/\psi$ associated with a gluon}},
  }{}\href{http://dx.doi.org/10.1016/0550-3213(80)90175-3}{Nucl.\ Phys.\  {\bf
  B172} (1980) 425}\relax
\mciteBstWouldAddEndPuncttrue
\mciteSetBstMidEndSepPunct{\mcitedefaultmidpunct}
{\mcitedefaultendpunct}{\mcitedefaultseppunct}\relax
\EndOfBibitem
\bibitem{Baier:1983va}
R.~Baier and R.~Ruckl, \ifthenelse{\boolean{articletitles}}{{\it {Hadronic
  collisions: A quarkonium factory}},
  }{}\href{http://dx.doi.org/10.1007/BF01572254}{Z.\ Phys.\  {\bf C19} (1983)
  251}\relax
\mciteBstWouldAddEndPuncttrue
\mciteSetBstMidEndSepPunct{\mcitedefaultmidpunct}
{\mcitedefaultendpunct}{\mcitedefaultseppunct}\relax
\EndOfBibitem
\bibitem{Bodwin:1994jh}
G.~T. Bodwin, E.~Braaten, and G.~P. Lepage,
  \ifthenelse{\boolean{articletitles}}{{\it {Rigorous QCD analysis of inclusive
  annihilation and production of heavy quarkonium}},
  }{}\href{http://dx.doi.org/10.1103/PhysRevD.51.1125}{Phys.\ Rev.\  {\bf D51}
  (1995) 1125}, \href{http://arxiv.org/abs/hep-ph/9407339}{{\tt
  arXiv:hep-ph/9407339}}, erratum ibid. {\bf D55} (1997) 5853\relax
\mciteBstWouldAddEndPuncttrue
\mciteSetBstMidEndSepPunct{\mcitedefaultmidpunct}
{\mcitedefaultendpunct}{\mcitedefaultseppunct}\relax
\EndOfBibitem
\bibitem{Cho:1995vh}
P.~L. Cho and A.~K. Leibovich, \ifthenelse{\boolean{articletitles}}{{\it {Color
  octet quarkonia production}},
  }{}\href{http://dx.doi.org/10.1103/PhysRevD.53.150}{Phys.\ Rev.\  {\bf D53}
  (1996) 150}, \href{http://arxiv.org/abs/hep-ph/9505329}{{\tt
  arXiv:hep-ph/9505329}}\relax
\mciteBstWouldAddEndPuncttrue
\mciteSetBstMidEndSepPunct{\mcitedefaultmidpunct}
{\mcitedefaultendpunct}{\mcitedefaultseppunct}\relax
\EndOfBibitem
\bibitem{Cho:1995ce}
P.~L. Cho and A.~K. Leibovich, \ifthenelse{\boolean{articletitles}}{{\it {Color
  octet quarkonia production. II}},
  }{}\href{http://dx.doi.org/10.1103/PhysRevD.53.6203}{Phys.\ Rev.\  {\bf D53}
  (1996) 6203}, \href{http://arxiv.org/abs/hep-ph/9511315}{{\tt
  arXiv:hep-ph/9511315}}\relax
\mciteBstWouldAddEndPuncttrue
\mciteSetBstMidEndSepPunct{\mcitedefaultmidpunct}
{\mcitedefaultendpunct}{\mcitedefaultseppunct}\relax
\EndOfBibitem
\bibitem{Abe:1997jz}
CDF collaboration, F.~Abe {\em et~al.},
  \ifthenelse{\boolean{articletitles}}{{\it {$J/\psi$ and $\psi(2S)$ production
  in $p\bar{p}$ collisions at $\sqrt{s} = 1.8$ TeV}},
  }{}\href{http://dx.doi.org/10.1103/PhysRevLett.79.572}{Phys.\ Rev.\ Lett.\
  {\bf 79} (1997) 572}\relax
\mciteBstWouldAddEndPuncttrue
\mciteSetBstMidEndSepPunct{\mcitedefaultmidpunct}
{\mcitedefaultendpunct}{\mcitedefaultseppunct}\relax
\EndOfBibitem
\bibitem{Abe:1997yz}
CDF collaboration, F.~Abe {\em et~al.},
  \ifthenelse{\boolean{articletitles}}{{\it {Production of $J/\psi$ mesons from
  $\chi_c$ meson decays in $p\bar{p}$ collisions at $\sqrt{s} = 1.8$ TeV}},
  }{}\href{http://dx.doi.org/10.1103/PhysRevLett.79.578}{Phys.\ Rev.\ Lett.\
  {\bf 79} (1997) 578}\relax
\mciteBstWouldAddEndPuncttrue
\mciteSetBstMidEndSepPunct{\mcitedefaultmidpunct}
{\mcitedefaultendpunct}{\mcitedefaultseppunct}\relax
\EndOfBibitem
\bibitem{Abulencia:2007us}
CDF collaboration, A.~Abulencia {\em et~al.},
  \ifthenelse{\boolean{articletitles}}{{\it {Polarization of $J/\psi$ and
  $\psi(2S)$ mesons produced in $p \bar{p}$ collisions at $\sqrt{s}$ = 1.96
  TeV}}, }{}\href{http://dx.doi.org/10.1103/PhysRevLett.99.132001}{Phys.\ Rev.\
  Lett.\  {\bf 99} (2007) 132001}, \href{http://arxiv.org/abs/0704.0638}{{\tt
  arXiv:0704.0638}}\relax
\mciteBstWouldAddEndPuncttrue
\mciteSetBstMidEndSepPunct{\mcitedefaultmidpunct}
{\mcitedefaultendpunct}{\mcitedefaultseppunct}\relax
\EndOfBibitem
\bibitem{Abelev:2011md}
ALICE collaboration, B.~Abelev {\em et~al.},
  \ifthenelse{\boolean{articletitles}}{{\it {$J/\psi$ polarization in $pp$
  collisions at $\sqrt{s}=7$ TeV}},
  }{}\href{http://dx.doi.org/10.1103/PhysRevLett.108.082001}{Phys.\ Rev.\
  Lett.\  {\bf 108} (2012) 082001}, \href{http://arxiv.org/abs/1111.1630}{{\tt
  arXiv:1111.1630}}\relax
\mciteBstWouldAddEndPuncttrue
\mciteSetBstMidEndSepPunct{\mcitedefaultmidpunct}
{\mcitedefaultendpunct}{\mcitedefaultseppunct}\relax
\EndOfBibitem
\bibitem{Chatrchyan:2012woa}
CMS collaboration, S.~Chatrchyan {\em et~al.},
  \ifthenelse{\boolean{articletitles}}{{\it {Measurement of the $\Upsilon(1S)$,
  $\Upsilon(2S)$ and $\Upsilon(3S)$ polarizations in $pp$ collisions at
  $\sqrt{s}=7$ TeV}},
  }{}\href{http://dx.doi.org/10.1103/PhysRevLett.110.081802}{Phys.\ Rev.\
  Lett.\  {\bf 110} (2013) 081802}, \href{http://arxiv.org/abs/1209.2922}{{\tt
  arXiv:1209.2922}}\relax
\mciteBstWouldAddEndPuncttrue
\mciteSetBstMidEndSepPunct{\mcitedefaultmidpunct}
{\mcitedefaultendpunct}{\mcitedefaultseppunct}\relax
\EndOfBibitem
\bibitem{Aaij:2013nlm}
LHCb collaboration, R.~Aaij {\em et~al.},
  \ifthenelse{\boolean{articletitles}}{{\it {Measurement of $J/\psi$
  polarization in $pp$ collisions at $\sqrt{s}=7$ TeV}},
  }{}\href{http://dx.doi.org/10.1140/epjc/s10052-013-2631-3}{Eur.\ Phys.\ J.\
  {\bf C73} (2013) 2631}, \href{http://arxiv.org/abs/1307.6379}{{\tt
  arXiv:1307.6379}}\relax
\mciteBstWouldAddEndPuncttrue
\mciteSetBstMidEndSepPunct{\mcitedefaultmidpunct}
{\mcitedefaultendpunct}{\mcitedefaultseppunct}\relax
\EndOfBibitem
\bibitem{Aaij:2014qea}
LHCb collaboration, R.~Aaij {\em et~al.},
  \ifthenelse{\boolean{articletitles}}{{\it {Measurement of $\psi(2S)$
  polarisation in $pp$ collisions at $\sqrt{s}=7$ TeV}},
  }{}\href{http://dx.doi.org/10.1140/epjc/s10052-014-2872-9}{Eur.\ Phys.\ J.\
  {\bf C74} (2014) 2872}, \href{http://arxiv.org/abs/1403.1339}{{\tt
  arXiv:1403.1339}}\relax
\mciteBstWouldAddEndPuncttrue
\mciteSetBstMidEndSepPunct{\mcitedefaultmidpunct}
{\mcitedefaultendpunct}{\mcitedefaultseppunct}\relax
\EndOfBibitem
\bibitem{Braaten:1993jn}
E.~Braaten, K.~Cheung, and T.~C. Yuan,
  \ifthenelse{\boolean{articletitles}}{{\it {Perturbative QCD fragmentation
  functions for $B_c$ and $B_{c}^{*}$ production}},
  }{}\href{http://dx.doi.org/10.1103/PhysRevD.48.R5049}{Phys.\ Rev.\  {\bf D48}
  (1993) R5049 (R)}, \href{http://arxiv.org/abs/hep-ph/9305206}{{\tt
  arXiv:hep-ph/9305206}}\relax
\mciteBstWouldAddEndPuncttrue
\mciteSetBstMidEndSepPunct{\mcitedefaultmidpunct}
{\mcitedefaultendpunct}{\mcitedefaultseppunct}\relax
\EndOfBibitem
\bibitem{Cheung:1999ir}
K.~Cheung, \ifthenelse{\boolean{articletitles}}{{\it {$B_c$ meson production at
  the Tevatron revisited}},
  }{}\href{http://dx.doi.org/10.1016/S0370-2693(99)01402-1}{Phys.\ Lett.\  {\bf
  B472} (2000) 408}, \href{http://arxiv.org/abs/hep-ph/9908405}{{\tt
  arXiv:hep-ph/9908405}}\relax
\mciteBstWouldAddEndPuncttrue
\mciteSetBstMidEndSepPunct{\mcitedefaultmidpunct}
{\mcitedefaultendpunct}{\mcitedefaultseppunct}\relax
\EndOfBibitem
\bibitem{Chang:1992jb}
C.-H. Chang and Y.-Q. Chen, \ifthenelse{\boolean{articletitles}}{{\it {Hadronic
  production of the $B_c$ meson at TeV energies}},
  }{}\href{http://dx.doi.org/10.1103/PhysRevD.48.4086}{Phys.\ Rev.\  {\bf D48}
  (1993) 4086}\relax
\mciteBstWouldAddEndPuncttrue
\mciteSetBstMidEndSepPunct{\mcitedefaultmidpunct}
{\mcitedefaultendpunct}{\mcitedefaultseppunct}\relax
\EndOfBibitem
\bibitem{Chang:1994aw}
C.-H. Chang, Y.-Q. Chen, G.-P. Han, and H.-T. Jiang,
  \ifthenelse{\boolean{articletitles}}{{\it {On hadronic production of the
  $B_c$ meson}},
  }{}\href{http://dx.doi.org/10.1016/0370-2693(95)01235-4}{Phys.\ Lett.\  {\bf
  B364} (1995) 78}, \href{http://arxiv.org/abs/hep-ph/9408242}{{\tt
  arXiv:hep-ph/9408242}}\relax
\mciteBstWouldAddEndPuncttrue
\mciteSetBstMidEndSepPunct{\mcitedefaultmidpunct}
{\mcitedefaultendpunct}{\mcitedefaultseppunct}\relax
\EndOfBibitem
\bibitem{Chang:1996jt}
C.-H. Chang, Y.-Q. Chen, and R.~J. Oakes,
  \ifthenelse{\boolean{articletitles}}{{\it {Comparative study of the hadronic
  production of $B_c$ mesons}},
  }{}\href{http://dx.doi.org/10.1103/PhysRevD.54.4344}{Phys.\ Rev.\  {\bf D54}
  (1996) 4344}, \href{http://arxiv.org/abs/hep-ph/9602411}{{\tt
  arXiv:hep-ph/9602411}}\relax
\mciteBstWouldAddEndPuncttrue
\mciteSetBstMidEndSepPunct{\mcitedefaultmidpunct}
{\mcitedefaultendpunct}{\mcitedefaultseppunct}\relax
\EndOfBibitem
\bibitem{Chang:2005bf}
C.-H. Chang, C.-F. Qiao, J.-X. Wang, and X.-G. Wu,
  \ifthenelse{\boolean{articletitles}}{{\it {Color-octet contributions to
  $P$-wave $B_c$ meson hadroproduction}},
  }{}\href{http://dx.doi.org/10.1103/PhysRevD.71.074012}{Phys.\ Rev.\  {\bf
  D71} (2005) 074012}, \href{http://arxiv.org/abs/hep-ph/0502155}{{\tt
  arXiv:hep-ph/0502155}}\relax
\mciteBstWouldAddEndPuncttrue
\mciteSetBstMidEndSepPunct{\mcitedefaultmidpunct}
{\mcitedefaultendpunct}{\mcitedefaultseppunct}\relax
\EndOfBibitem
\bibitem{Kolodziej:1995nv}
K.~Kolodziej, A.~Leike, and R.~Ruckl, \ifthenelse{\boolean{articletitles}}{{\it
  {Production of $B_c$ mesons in hadronic collisions}},
  }{}\href{http://dx.doi.org/10.1016/0370-2693(95)00710-3}{Phys.\ Lett.\  {\bf
  B355} (1995) 337}, \href{http://arxiv.org/abs/hep-ph/9505298}{{\tt
  arXiv:hep-ph/9505298}}\relax
\mciteBstWouldAddEndPuncttrue
\mciteSetBstMidEndSepPunct{\mcitedefaultmidpunct}
{\mcitedefaultendpunct}{\mcitedefaultseppunct}\relax
\EndOfBibitem
\bibitem{Berezhnoy:1994ba}
A.~V. Berezhnoy, A.~K. Likhoded, and M.~V. Shevlyagin,
  \ifthenelse{\boolean{articletitles}}{{\it {Hadronic production of $B_c$
  mesons}}, }{}Phys.\ Atom.\ Nucl.\  {\bf 58} (1995) 672,
  \href{http://arxiv.org/abs/hep-ph/9408284}{{\tt arXiv:hep-ph/9408284}}\relax
\mciteBstWouldAddEndPuncttrue
\mciteSetBstMidEndSepPunct{\mcitedefaultmidpunct}
{\mcitedefaultendpunct}{\mcitedefaultseppunct}\relax
\EndOfBibitem
\bibitem{Berezhnoy:1996an}
A.~V. Berezhnoy, V.~V. Kiselev, and A.~K. Likhoded,
  \ifthenelse{\boolean{articletitles}}{{\it {Photonic production of $S$- and
  $P$-wave $B_c$ states and doubly heavy baryons}},
  }{}\href{http://dx.doi.org/10.1007/s002180050152}{Z.\ Phys.\  {\bf A356}
  (1996) 89}\relax
\mciteBstWouldAddEndPuncttrue
\mciteSetBstMidEndSepPunct{\mcitedefaultmidpunct}
{\mcitedefaultendpunct}{\mcitedefaultseppunct}\relax
\EndOfBibitem
\bibitem{Baranov:1997wy}
S.~P. Baranov, \ifthenelse{\boolean{articletitles}}{{\it {Pair production of
  $B_c^{(*)}$ mesons in $pp$ and $\gamma\gamma$ collisions}},
  }{}\href{http://dx.doi.org/10.1103/PhysRevD.55.2756}{Phys.\ Rev.\  {\bf D55}
  (1997) 2756}\relax
\mciteBstWouldAddEndPuncttrue
\mciteSetBstMidEndSepPunct{\mcitedefaultmidpunct}
{\mcitedefaultendpunct}{\mcitedefaultseppunct}\relax
\EndOfBibitem
\bibitem{Chang:2003cr}
C.-H. Chang and X.-G. Wu, \ifthenelse{\boolean{articletitles}}{{\it
  {Uncertainties in estimating $B_c$ hadronic production and comparisons of the
  production at TEVATRON and LHC}},
  }{}\href{http://dx.doi.org/10.1140/epjc/s2004-02015-0}{Eur.\ Phys.\ J.\  {\bf
  C38} (2004) 267}, \href{http://arxiv.org/abs/hep-ph/0309121}{{\tt
  arXiv:hep-ph/0309121}}\relax
\mciteBstWouldAddEndPuncttrue
\mciteSetBstMidEndSepPunct{\mcitedefaultmidpunct}
{\mcitedefaultendpunct}{\mcitedefaultseppunct}\relax
\EndOfBibitem
\bibitem{Gao:2010zzc}
Y.-N. Gao {\em et~al.}, \ifthenelse{\boolean{articletitles}}{{\it {Experimental
  prospects of the $B_c$ studies of the LHCb experiment}},
  }{}\href{http://dx.doi.org/10.1088/0256-307X/27/6/061302}{Chin.\ Phys.\
  Lett.\  {\bf 27} (2010) 061302}\relax
\mciteBstWouldAddEndPuncttrue
\mciteSetBstMidEndSepPunct{\mcitedefaultmidpunct}
{\mcitedefaultendpunct}{\mcitedefaultseppunct}\relax
\EndOfBibitem
\bibitem{LHCb-PAPER-2011-003}
LHCb collaboration, R.~Aaij {\em et~al.},
  \ifthenelse{\boolean{articletitles}}{{\it {Measurement of $J/\psi$ production
  in $pp$ collisions at $\sqrt{s}=7$ TeV}},
  }{}\href{http://dx.doi.org/10.1140/epjc/s10052-011-1645-y}{Eur.\ Phys.\ J.\
  {\bf C71} (2011) 1645}, \href{http://arxiv.org/abs/1103.0423}{{\tt
  arXiv:1103.0423}}\relax
\mciteBstWouldAddEndPuncttrue
\mciteSetBstMidEndSepPunct{\mcitedefaultmidpunct}
{\mcitedefaultendpunct}{\mcitedefaultseppunct}\relax
\EndOfBibitem
\bibitem{Abe:1998wi}
CDF collaboration, F.~Abe {\em et~al.},
  \ifthenelse{\boolean{articletitles}}{{\it {Observation of the $B_c$ meson in
  $p\bar{p}$ collisions at $\sqrt{s} = 1.8$ TeV}},
  }{}\href{http://dx.doi.org/10.1103/PhysRevLett.81.2432}{Phys.\ Rev.\ Lett.\
  {\bf 81} (1998) 2432}, \href{http://arxiv.org/abs/hep-ex/9805034}{{\tt
  arXiv:hep-ex/9805034}}\relax
\mciteBstWouldAddEndPuncttrue
\mciteSetBstMidEndSepPunct{\mcitedefaultmidpunct}
{\mcitedefaultendpunct}{\mcitedefaultseppunct}\relax
\EndOfBibitem
\bibitem{Abe:1998fb}
CDF collaboration, F.~Abe {\em et~al.},
  \ifthenelse{\boolean{articletitles}}{{\it {Observation of $B_c$ mesons in
  $p\bar{p}$ collisions at $\sqrt{s} = 1.8$ TeV}},
  }{}\href{http://dx.doi.org/10.1103/PhysRevD.58.112004}{Phys.\ Rev.\  {\bf
  D58} (1998) 112004}, \href{http://arxiv.org/abs/hep-ex/9804014}{{\tt
  arXiv:hep-ex/9804014}}\relax
\mciteBstWouldAddEndPuncttrue
\mciteSetBstMidEndSepPunct{\mcitedefaultmidpunct}
{\mcitedefaultendpunct}{\mcitedefaultseppunct}\relax
\EndOfBibitem
\bibitem{LHCB-PAPER-2012-028}
LHCb collaboration, R.~Aaij {\em et~al.},
  \ifthenelse{\boolean{articletitles}}{{\it {Measurements of $B_c^+$ production
  and mass with the $B_c^+ \to J/\psi \pi^+$ decay}},
  }{}\href{http://dx.doi.org/10.1103/PhysRevLett.109.232001}{Phys.\ Rev.\
  Lett.\  {\bf 109} (2012) 232001}, \href{http://arxiv.org/abs/1209.5634}{{\tt
  arXiv:1209.5634}}\relax
\mciteBstWouldAddEndPuncttrue
\mciteSetBstMidEndSepPunct{\mcitedefaultmidpunct}
{\mcitedefaultendpunct}{\mcitedefaultseppunct}\relax
\EndOfBibitem
\bibitem{LHCb-PAPER-2013-044}
LHCb collaboration, R.~Aaij {\em et~al.},
  \ifthenelse{\boolean{articletitles}}{{\it {Observation of the decay $B_c^+\to
  B_s^0\pi^+$}},
  }{}\href{http://dx.doi.org/10.1103/PhysRevLett.111.181801}{Phys.\ Rev.\
  Lett.\  {\bf 111} (2013) 181801}, \href{http://arxiv.org/abs/1308.4544}{{\tt
  arXiv:1308.4544}}\relax
\mciteBstWouldAddEndPuncttrue
\mciteSetBstMidEndSepPunct{\mcitedefaultmidpunct}
{\mcitedefaultendpunct}{\mcitedefaultseppunct}\relax
\EndOfBibitem
\bibitem{Aaltonen:2008eu}
CDF collaboration, T.~Aaltonen {\em et~al.},
  \ifthenelse{\boolean{articletitles}}{{\it {First measurement of the ratio of
  branching fractions $B(\Lambda^0_b \to \Lambda^+_{c} \mu^{-} \bar{\nu}_\mu) /
  B(\Lambda^0_b \to \Lambda^+_{c} \pi^{-})$}},
  }{}\href{http://dx.doi.org/10.1103/PhysRevD.79.032001}{Phys.\ Rev.\  {\bf
  D79} (2009) 032001}, \href{http://arxiv.org/abs/0810.3213}{{\tt
  arXiv:0810.3213}}\relax
\mciteBstWouldAddEndPuncttrue
\mciteSetBstMidEndSepPunct{\mcitedefaultmidpunct}
{\mcitedefaultendpunct}{\mcitedefaultseppunct}\relax
\EndOfBibitem
\bibitem{LHCb-PAPER-2011-018}
LHCb collaboration, R.~Aaij {\em et~al.},
  \ifthenelse{\boolean{articletitles}}{{\it {Measurement of $b$ hadron
  production fractions in 7 TeV $pp$ collisions}},
  }{}\href{http://dx.doi.org/10.1103/PhysRevD.85.032008}{Phys.\ Rev.\  {\bf
  D85} (2012) 032008}, \href{http://arxiv.org/abs/1111.2357}{{\tt
  arXiv:1111.2357}}\relax
\mciteBstWouldAddEndPuncttrue
\mciteSetBstMidEndSepPunct{\mcitedefaultmidpunct}
{\mcitedefaultendpunct}{\mcitedefaultseppunct}\relax
\EndOfBibitem
\bibitem{LHCB-PAPER-2014-004}
LHCb collaboration, R.~Aaij {\em et~al.},
  \ifthenelse{\boolean{articletitles}}{{\it {Study of the kinematic dependences
  of $\Lambda_b^0$ production in $pp$ collisions and a measurement of the
  $\Lambda_b^0 \to \Lambda_c^+ \pi^-$ branching fraction}},
  }{}\href{http://dx.doi.org/10.1007/JHEP08(2014)143}{JHEP {\bf 08} (2014)
  143}, \href{http://arxiv.org/abs/1405.6842}{{\tt arXiv:1405.6842}}\relax
\mciteBstWouldAddEndPuncttrue
\mciteSetBstMidEndSepPunct{\mcitedefaultmidpunct}
{\mcitedefaultendpunct}{\mcitedefaultseppunct}\relax
\EndOfBibitem
\bibitem{LHCB-PAPER-2014-002}
LHCb collaboration, R.~Aaij {\em et~al.},
  \ifthenelse{\boolean{articletitles}}{{\it {Study of beauty hadron decays into
  pairs of charm hadrons}},
  }{}\href{http://dx.doi.org/10.1103/PhysRevLett.112.202001}{Phys.\ Rev.\
  Lett.\  {\bf 112} (2014) 202001}, \href{http://arxiv.org/abs/1403.3606}{{\tt
  arXiv:1403.3606}}\relax
\mciteBstWouldAddEndPuncttrue
\mciteSetBstMidEndSepPunct{\mcitedefaultmidpunct}
{\mcitedefaultendpunct}{\mcitedefaultseppunct}\relax
\EndOfBibitem
\bibitem{Alves:2008zz}
LHCb collaboration, A.~A. Alves~Jr. {\em et~al.},
  \ifthenelse{\boolean{articletitles}}{{\it {The \lhcb detector at the LHC}},
  }{}\href{http://dx.doi.org/10.1088/1748-0221/3/08/S08005}{JINST {\bf 3}
  (2008) S08005}\relax
\mciteBstWouldAddEndPuncttrue
\mciteSetBstMidEndSepPunct{\mcitedefaultmidpunct}
{\mcitedefaultendpunct}{\mcitedefaultseppunct}\relax
\EndOfBibitem
\bibitem{LHCb-DP-2012-004}
R.~Aaij {\em et~al.}, \ifthenelse{\boolean{articletitles}}{{\it {The \lhcb
  trigger and its performance in 2011}},
  }{}\href{http://dx.doi.org/10.1088/1748-0221/8/04/P04022}{JINST {\bf 8}
  (2013) P04022}, \href{http://arxiv.org/abs/1211.3055}{{\tt
  arXiv:1211.3055}}\relax
\mciteBstWouldAddEndPuncttrue
\mciteSetBstMidEndSepPunct{\mcitedefaultmidpunct}
{\mcitedefaultendpunct}{\mcitedefaultseppunct}\relax
\EndOfBibitem
\bibitem{PDG2014}
Particle Data Group, K.~A. Olive {\em et~al.},
  \ifthenelse{\boolean{articletitles}}{{\it {\href{http://pdg.lbl.gov/}{The
  review of particle physics}}}, }{}Chin.\ Phys.\ C {\bf 38} (2014)
  090001\relax
\mciteBstWouldAddEndPuncttrue
\mciteSetBstMidEndSepPunct{\mcitedefaultmidpunct}
{\mcitedefaultendpunct}{\mcitedefaultseppunct}\relax
\EndOfBibitem
\bibitem{BBDT}
V.~V. Gligorov and M.~Williams, \ifthenelse{\boolean{articletitles}}{{\it
  {Efficient, reliable and fast high-level triggering using a bonsai boosted
  decision tree}},
  }{}\href{http://dx.doi.org/10.1088/1748-0221/8/02/P02013}{JINST {\bf 8}
  (2013) P02013}, \href{http://arxiv.org/abs/1210.6861}{{\tt
  arXiv:1210.6861}}\relax
\mciteBstWouldAddEndPuncttrue
\mciteSetBstMidEndSepPunct{\mcitedefaultmidpunct}
{\mcitedefaultendpunct}{\mcitedefaultseppunct}\relax
\EndOfBibitem
\bibitem{Breiman}
L.~Breiman, J.~H. Friedman, R.~A. Olshen, and C.~J. Stone, {\em Classification
  and regression trees}, Wadsworth international group, Belmont, California,
  USA, 1984\relax
\mciteBstWouldAddEndPuncttrue
\mciteSetBstMidEndSepPunct{\mcitedefaultmidpunct}
{\mcitedefaultendpunct}{\mcitedefaultseppunct}\relax
\EndOfBibitem
\bibitem{AdaBoost}
R.~E. Schapire and Y.~Freund, \ifthenelse{\boolean{articletitles}}{{\it A
  decision-theoretic generalization of on-line learning and an application to
  boosting}, }{}\href{http://dx.doi.org/10.1006/jcss.1997.1504}{Jour.\ Comp.\
  and Syst.\ Sc.\  {\bf 55} (1997) 119}\relax
\mciteBstWouldAddEndPuncttrue
\mciteSetBstMidEndSepPunct{\mcitedefaultmidpunct}
{\mcitedefaultendpunct}{\mcitedefaultseppunct}\relax
\EndOfBibitem
\bibitem{Sjostrand:2006za}
T.~Sj\"{o}strand, S.~Mrenna, and P.~Skands,
  \ifthenelse{\boolean{articletitles}}{{\it {PYTHIA 6.4 physics and manual}},
  }{}\href{http://dx.doi.org/10.1088/1126-6708/2006/05/026}{JHEP {\bf 05}
  (2006) 026}, \href{http://arxiv.org/abs/hep-ph/0603175}{{\tt
  arXiv:hep-ph/0603175}}\relax
\mciteBstWouldAddEndPuncttrue
\mciteSetBstMidEndSepPunct{\mcitedefaultmidpunct}
{\mcitedefaultendpunct}{\mcitedefaultseppunct}\relax
\EndOfBibitem
\bibitem{LHCb-PROC-2010-056}
I.~Belyaev {\em et~al.}, \ifthenelse{\boolean{articletitles}}{{\it {Handling of
  the generation of primary events in \gauss, the \lhcb simulation framework}},
  }{}\href{http://dx.doi.org/10.1109/NSSMIC.2010.5873949}{Nuclear Science
  Symposium Conference Record (NSS/MIC) {\bf IEEE} (2010) 1155}\relax
\mciteBstWouldAddEndPuncttrue
\mciteSetBstMidEndSepPunct{\mcitedefaultmidpunct}
{\mcitedefaultendpunct}{\mcitedefaultseppunct}\relax
\EndOfBibitem
\bibitem{Chang:2005hq}
C.-H. Chang, J.-X. Wang, and X.-G. Wu,
  \ifthenelse{\boolean{articletitles}}{{\it {BCVEGPY2.0: An upgraded version of
  the generator BCVEGPY with the addition of hadroproduction of the $P$-wave
  $B_c$ states}},
  }{}\href{http://dx.doi.org/10.1016/j.cpc.2005.09.008}{Comput.\ Phys.\
  Commun.\  {\bf 174} (2006) 241},
  \href{http://arxiv.org/abs/hep-ph/0504017}{{\tt arXiv:hep-ph/0504017}}\relax
\mciteBstWouldAddEndPuncttrue
\mciteSetBstMidEndSepPunct{\mcitedefaultmidpunct}
{\mcitedefaultendpunct}{\mcitedefaultseppunct}\relax
\EndOfBibitem
\bibitem{Lange:2001uf}
D.~J. Lange, \ifthenelse{\boolean{articletitles}}{{\it {The EvtGen particle
  decay simulation package}},
  }{}\href{http://dx.doi.org/10.1016/S0168-9002(01)00089-4}{Nucl.\ Instrum.\
  Meth.\  {\bf A462} (2001) 152}\relax
\mciteBstWouldAddEndPuncttrue
\mciteSetBstMidEndSepPunct{\mcitedefaultmidpunct}
{\mcitedefaultendpunct}{\mcitedefaultseppunct}\relax
\EndOfBibitem
\bibitem{Golonka:2005pn}
P.~Golonka and Z.~Was, \ifthenelse{\boolean{articletitles}}{{\it {PHOTOS Monte
  Carlo: A precision tool for QED corrections in $Z$ and $W$ decays}},
  }{}\href{http://dx.doi.org/10.1140/epjc/s2005-02396-4}{Eur.\ Phys.\ J.\  {\bf
  C45} (2006) 97}, \href{http://arxiv.org/abs/hep-ph/0506026}{{\tt
  arXiv:hep-ph/0506026}}\relax
\mciteBstWouldAddEndPuncttrue
\mciteSetBstMidEndSepPunct{\mcitedefaultmidpunct}
{\mcitedefaultendpunct}{\mcitedefaultseppunct}\relax
\EndOfBibitem
\bibitem{Allison:2006ve}
Geant4 collaboration, J.~Allison {\em et~al.},
  \ifthenelse{\boolean{articletitles}}{{\it {Geant4 developments and
  applications}}, }{}\href{http://dx.doi.org/10.1109/TNS.2006.869826}{IEEE
  Trans.\ Nucl.\ Sci.\  {\bf 53} (2006) 270}\relax
\mciteBstWouldAddEndPuncttrue
\mciteSetBstMidEndSepPunct{\mcitedefaultmidpunct}
{\mcitedefaultendpunct}{\mcitedefaultseppunct}\relax
\EndOfBibitem
\bibitem{Agostinelli:2002hh}
Geant4 collaboration, S.~Agostinelli {\em et~al.},
  \ifthenelse{\boolean{articletitles}}{{\it {Geant4: A simulation toolkit}},
  }{}\href{http://dx.doi.org/10.1016/S0168-9002(03)01368-8}{Nucl.\ Instrum.\
  Meth.\  {\bf A506} (2003) 250}\relax
\mciteBstWouldAddEndPuncttrue
\mciteSetBstMidEndSepPunct{\mcitedefaultmidpunct}
{\mcitedefaultendpunct}{\mcitedefaultseppunct}\relax
\EndOfBibitem
\bibitem{LHCb-PROC-2011-006}
M.~Clemencic {\em et~al.}, \ifthenelse{\boolean{articletitles}}{{\it {The \lhcb
  simulation application, \gauss: design, evolution and experience}},
  }{}\href{http://dx.doi.org/10.1088/1742-6596/331/3/032023}{{J.\ Phys.\ Conf.\
  Ser.\ } {\bf 331} (2011) 032023}\relax
\mciteBstWouldAddEndPuncttrue
\mciteSetBstMidEndSepPunct{\mcitedefaultmidpunct}
{\mcitedefaultendpunct}{\mcitedefaultseppunct}\relax
\EndOfBibitem
\bibitem{LHCb-PAPER-2013-063}
LHCb collaboration, R.~Aaij {\em et~al.},
  \ifthenelse{\boolean{articletitles}}{{\it {Measurement of the $B_c^+$ meson
  lifetime using $B_c^+\to J/\psi\mu^+\nu_{\mu} X$ decays}},
  }{}\href{http://dx.doi.org/10.1140/epjc/s10052-014-2839-x}{Eur.\ Phys.\ J.\
  {\bf C74} (2014) 2839}, \href{http://arxiv.org/abs/1401.6932}{{\tt
  arXiv:1401.6932}}\relax
\mciteBstWouldAddEndPuncttrue
\mciteSetBstMidEndSepPunct{\mcitedefaultmidpunct}
{\mcitedefaultendpunct}{\mcitedefaultseppunct}\relax
\EndOfBibitem
\bibitem{Hulsbergen:2005pu}
W.~D. Hulsbergen, \ifthenelse{\boolean{articletitles}}{{\it {Decay chain
  fitting with a Kalman filter}},
  }{}\href{http://dx.doi.org/10.1016/j.nima.2005.06.078}{Nucl.\ Instrum.\
  Meth.\  {\bf A552} (2005) 566},
  \href{http://arxiv.org/abs/physics/0503191}{{\tt
  arXiv:physics/0503191}}\relax
\mciteBstWouldAddEndPuncttrue
\mciteSetBstMidEndSepPunct{\mcitedefaultmidpunct}
{\mcitedefaultendpunct}{\mcitedefaultseppunct}\relax
\EndOfBibitem
\bibitem{LHCB-PAPER-2013-021}
LHCb collaboration, R.~Aaij {\em et~al.},
  \ifthenelse{\boolean{articletitles}}{{\it {First observation of the decay
  $B_c^+ \to J/\psi K^+$}},
  }{}\href{http://dx.doi.org/10.1007/JHEP09(2013)075}{JHEP {\bf 09} (2013)
  075}, \href{http://arxiv.org/abs/1306.6723}{{\tt arXiv:1306.6723}}\relax
\mciteBstWouldAddEndPuncttrue
\mciteSetBstMidEndSepPunct{\mcitedefaultmidpunct}
{\mcitedefaultendpunct}{\mcitedefaultseppunct}\relax
\EndOfBibitem
\bibitem{LHCB-PAPER-2011-024}
LHCb collaboration, R.~Aaij {\em et~al.},
  \ifthenelse{\boolean{articletitles}}{{\it {Measurements of the branching
  fractions and $CP$ asymmetries of $B^\pm \to J/\psi \pi^\pm$ and $B^\pm \to
  \psi (2S) \pi^\pm$ decays}},
  }{}\href{http://dx.doi.org/10.1103/PhysRevD.85.091105}{Phys.\ Rev.\  {\bf
  D85} (2012) 091105(R)}, \href{http://arxiv.org/abs/1203.3592}{{\tt
  arXiv:1203.3592}}\relax
\mciteBstWouldAddEndPuncttrue
\mciteSetBstMidEndSepPunct{\mcitedefaultmidpunct}
{\mcitedefaultendpunct}{\mcitedefaultseppunct}\relax
\EndOfBibitem
\bibitem{suppmat}
See supplementary material for details, which includes Refs. [52-55]\relax
\mciteBstWouldAddEndPuncttrue
\mciteSetBstMidEndSepPunct{\mcitedefaultmidpunct}
{\mcitedefaultendpunct}{\mcitedefaultseppunct}\relax
\EndOfBibitem
\bibitem{Cacciari:2012ny}
M.~Cacciari {\em et~al.}, \ifthenelse{\boolean{articletitles}}{{\it
  {Theoretical predictions for charm and bottom production at the LHC}},
  }{}\href{http://dx.doi.org/10.1007/JHEP10(2012)137}{JHEP {\bf 10} (2012)
  137}, \href{http://arxiv.org/abs/1205.6344}{{\tt arXiv:1205.6344}}\relax
\mciteBstWouldAddEndPuncttrue
\mciteSetBstMidEndSepPunct{\mcitedefaultmidpunct}
{\mcitedefaultendpunct}{\mcitedefaultseppunct}\relax
\EndOfBibitem
\bibitem{Qiao:2012hp}
C.-F. Qiao, P.~Sun, D.~Yang, and R.-L. Zhu,
  \ifthenelse{\boolean{articletitles}}{{\it {$B_c$ exclusive decays to
  charmonium and a light meson at next-to-leading order accuracy}},
  }{}\href{http://dx.doi.org/10.1103/PhysRevD.89.034008}{Phys.\ Rev.\  {\bf
  D89} (2014) 034008}, \href{http://arxiv.org/abs/1209.5859}{{\tt
  arXiv:1209.5859}}\relax
\mciteBstWouldAddEndPuncttrue
\mciteSetBstMidEndSepPunct{\mcitedefaultmidpunct}
{\mcitedefaultendpunct}{\mcitedefaultseppunct}\relax
\EndOfBibitem
\bibitem{Pumplin:2002vw}
J.~Pumplin {\em et~al.}, \ifthenelse{\boolean{articletitles}}{{\it {New
  generation of parton distributions with uncertainties from global QCD
  analysis}}, }{}\href{http://dx.doi.org/10.1088/1126-6708/2002/07/012}{JHEP
  {\bf 07} (2002) 012}, \href{http://arxiv.org/abs/hep-ph/0201195}{{\tt
  arXiv:hep-ph/0201195}}\relax
\mciteBstWouldAddEndPuncttrue
\mciteSetBstMidEndSepPunct{\mcitedefaultmidpunct}
{\mcitedefaultendpunct}{\mcitedefaultseppunct}\relax
\EndOfBibitem
\bibitem{Nadolsky:2008zw}
P.~M. Nadolsky {\em et~al.}, \ifthenelse{\boolean{articletitles}}{{\it
  {Implications of CTEQ global analysis for collider observables}},
  }{}\href{http://dx.doi.org/10.1103/PhysRevD.78.013004}{Phys.\ Rev.\  {\bf
  D78} (2008) 013004}, \href{http://arxiv.org/abs/0802.0007}{{\tt
  arXiv:0802.0007}}\relax
\mciteBstWouldAddEndPuncttrue
\mciteSetBstMidEndSepPunct{\mcitedefaultmidpunct}
{\mcitedefaultendpunct}{\mcitedefaultseppunct}\relax
\EndOfBibitem
\bibitem{LHCb-PAPER-2013-004}
LHCb collaboration, R.~Aaij {\em et~al.},
  \ifthenelse{\boolean{articletitles}}{{\it {Measurement of $B$ meson
  production cross-sections in proton-proton collisions at $\sqrt{s} = 7$
  TeV}}, }{}\href{http://dx.doi.org/10.1007/JHEP08(2013)117}{JHEP {\bf 08}
  (2013) 117}, \href{http://arxiv.org/abs/1306.3663}{{\tt
  arXiv:1306.3663}}\relax
\mciteBstWouldAddEndPuncttrue
\mciteSetBstMidEndSepPunct{\mcitedefaultmidpunct}
{\mcitedefaultendpunct}{\mcitedefaultseppunct}\relax
\EndOfBibitem
\bibitem{LHCb-PAPER-2013-016}
LHCb collaboration, R.~Aaij {\em et~al.},
  \ifthenelse{\boolean{articletitles}}{{\it {Production of $J/\psi$ and
  $\Upsilon$ mesons in $pp$ collisions at $\sqrt{s}=8$ TeV}},
  }{}\href{http://dx.doi.org/10.1007/JHEP06(2013)064}{JHEP {\bf 06} (2013)
  064}, \href{http://arxiv.org/abs/1304.6977}{{\tt arXiv:1304.6977}}\relax
\mciteBstWouldAddEndPuncttrue
\mciteSetBstMidEndSepPunct{\mcitedefaultmidpunct}
{\mcitedefaultendpunct}{\mcitedefaultseppunct}\relax
\EndOfBibitem
\end{mcitethebibliography}

\newpage

%%%%%%%%%%%%%%%%%%%%%%%%%%%%%%%%%%%%%%%%%%
\centerline{\large\bf LHCb collaboration}
\begin{flushleft}
\small
R.~Aaij$^{41}$,
B.~Adeva$^{37}$,
M.~Adinolfi$^{46}$,
A.~Affolder$^{52}$,
Z.~Ajaltouni$^{5}$,
S.~Akar$^{6}$,
J.~Albrecht$^{9}$,
F.~Alessio$^{38}$,
M.~Alexander$^{51}$,
S.~Ali$^{41}$
% ...

{\footnotesize \it
% ...
$ ^{5}$Clermont Universit\'{e}, Universit\'{e} Blaise Pascal, CNRS/IN2P3, LPC, Clermont-Ferrand, France\\
$ ^{6}$CPPM, Aix-Marseille Universit\'{e}, CNRS/IN2P3, Marseille, France\\
% ...
$ ^{9}$Fakult\"{a}t Physik, Technische Universit\"{a}t Dortmund, Dortmund, Germany\\
% ...
$ ^{37}$Universidad de Santiago de Compostela, Santiago de Compostela, Spain\\
$ ^{38}$European Organization for Nuclear Research (CERN), Geneva, Switzerland\\
% ...
$ ^{41}$Nikhef National Institute for Subatomic Physics, Amsterdam, The Netherlands\\
% ...
$ ^{46}$H.H. Wills Physics Laboratory, University of Bristol, Bristol, United Kingdom\\
% ...
$ ^{51}$School of Physics and Astronomy, University of Glasgow, Glasgow, United Kingdom\\
$ ^{52}$Oliver Lodge Laboratory, University of Liverpool, Liverpool, United Kingdom\\
% ...
}
\end{flushleft}
%%%%%%%%%%%%%%%%%%%%%%%%%%%%%%%%%%%%%%%%%%

\end{document}